\documentclass{aa}  

\usepackage{graphicx}
\usepackage{txfonts}
\usepackage{lscape}
\usepackage[colorlinks=true, linkcolor=blue, citecolor=blue, filecolor=blue, urlcolor=blue]{hyperref}
\begin{document}

   \title{The SOFIA Massive (SOMA) Star Formation Q-band Follow-up. I.}

   \subtitle{Carbon-chain chemistry of intermediate-mass protostars}

   \author{Kotomi Taniguchi \inst{1}, Prasanta Gorai \inst{2,3}, Jonathan C. Tan \inst{4,5}, Miguel G$\acute{\rm {o}}$mez-Garrido\inst{6}, Rub{\'e}n Fedriani\inst{7}, Yao-Lun Yang\inst{8}, Tirupati Kumara Sridharan\inst{9}, Kei E. I. Tanaka\inst{10}, Masao Saito\inst{1,11}, Yichen Zhang\inst{12}, Lawrence Morgan\inst{13}, Giuliana Cosentino\inst{14}, Chi-Yan Law\inst{15}
          }

   \institute{National Astronomical Observatory of Japan, National Institutes of Natural Sciences,
              2-21-1 Osawa, Mitaka, Tokyo, 181-8588, Japan \\
              \email{kotomi.taniguchi@nao.ac.jp}
              \and
              Rosseland Centre for Solar Physics, University of Oslo, PO Box 1029 Blindern, 0315, Oslo, Norway
              \and
              Institute of Theoretical Astrophysics, University of Oslo, PO Box 1029 Blindern, 0315, Oslo, Norway
              \and
              Department of Astronomy, University of Virginia, Charlottesville, VA 22904, USA
              \and
              Department of Space, Earth \& Environment, Chalmers University of Technology, 412 93  Gothenburg, Sweden
              \and
              Observatorio Astronomico Nacional (OAN-IGN), Alfonso XII 3, 28014, Madrid, Apain
              \and
              Instituto de Astrof\'isica de Andaluc\'ia, CSIC, Glorieta de la Astronom\'ia s/n, E-18008 Granada, Spain
              \and
              Star and Planet Formation Laboratory, RIKEN Cluster for Pioneering Research, Wako, Saitama 351-0198, Japan
              \and
              National Radio Astronomy Observatory, 520 Edgemont Rd., Charlottesville, VA 22903, USA
              \and
              Department of Earth and Planetary Sciences, Institute of Science Tokyo, Meguro, Tokyo, 152-8551, Japan
              \and
              Graduate Institute for Advanced Studies, SOKENDAI, 2-21-1 Osawa, Mitaka, Tokyo 181-8588, Japan
              \and
              Department of Astronomy, Shanghai Jiao Tong University, 800 Dongchuan Rd., Minhang, Shanghai 200240, People’s Republic of China
              \and
              Green Bank Observatory, 155 Observatory Rd, Green Bank, WV 24944, USA
              \and
              European Southern Observatory, Karl-Schwarzschild-Str. 285748 Garching bei, M\"{u}nchen, Germany
              \and
              Osservatorio Astrofisico di Arcetri, Largo Enrico Fermi, 5, 50125 Firenze FI, Italy
             }

   \date{Received xxx, 2024; accepted xxx}

 
  \abstract
   {Evidence for similar chemical characteristics around low- and high-mass protostars has been found: in particular, a variety of carbon-chain species and complex organic molecules (COMs) are formed around them. On the other hand, the chemical compositions around intermediate-mass (IM; $2 M_{\odot} < m_* <8 M_{\odot}$) protostars have not been studied with large samples. In particular, it is unclear the extent to which carbon-chain species are formed around them.}
   {We aim to obtain the chemical compositions, particularly focusing on carbon-chain species, towards a sample of IM protostars. Another purpose is deriving the rotational temperatures of HC$_5$N to confirm whether carbon-chain species are formed in the warm gas around these stars.}
   {We have conducted Q-band (31.5--50 GHz) line survey observations towards eleven mainly intermediate-mass protostars with the Yebes 40 m radio telescope. The target protostars were selected from a sub-sample of the source list of the SOFIA Massive (SOMA) Star Formation project. Assuming local thermodynamic equilibrium (LTE), we have derived column densities of the detected molecules and rotational temperatures of HC$_5$N and CH$_3$OH.}
   {Nine carbon-chain species (HC$_3$N, HC$_5$N, C$_3$H, C$_4$H, $linear-$H$_2$CCC, $cyclic-$C$_3$H$_2$, CCS, C$_3$S, and CH$_3$CCH), three COMs (CH$_3$OH, CH$_3$CHO, and CH$_3$CN), H$_2$CCO, HNCO, and four simple sulfur (S)-bearing species ($^{13}$CS, C$^{34}$S, HCS$^+$, H$_2$CS) have been detected. 
   The rotational temperatures of HC$_5$N are derived to be $\sim20-30$ K in three IM protostars (Cepheus E, HH288, and IRAS\,20293+3952). The rotational temperatures of CH$_3$OH are derived in five IM sources and found to be similar to those of HC$_5$N.}
   {The rotational temperatures of HC$_5$N around the three IM protostars are 
   very similar compared to those around low- and high-mass protostars. These results indicate that carbon-chain molecules are formed in lukewarm ($\sim20-30$ K) gas around the IM protostars by the Warm Carbon-Chain Chemistry (WCCC) process. Thus, carbon-chain formation occurs ubiquitously in the warm gas around protostars across a wide range of stellar masses. Carbon-chain molecules and COMs coexist around most of the target IM protostars, which is similar to the situation in low- and high-mass protostars. In summary, the chemical characteristics around protostars are common in the low-, intermediate- and high-mass regimes.}


\authorrunning{Taniguchi, Gorai, Tan, et al.}
\maketitle


\nolinenumbers
\section{Introduction} \label{sec:intro}

Many astrochemical studies have been dedicated to investigating the chemical compositions around protostars \citep[for a review see][]{2020ARA&A..58..727J}.
It has been well-known that complex organic molecules (COMs), which consist of more than six atoms \citep{2009ARA&A..47..427H}, are abundant in hot regions with temperatures $\geq$ 100 K, namely hot cores and hot corinos around high-mass ($m_* \geq 8 M_{\odot}$) and low-mass ($m_* \leq 2 M_{\odot}$) protostars, respectively.
These COMs are formed on dust surfaces during the cold prestellar core stage and/or the warm-up stage after protostars are born, or synthesized in hot gas around protostars \citep[e.g.,][]{2019MNRAS.482.3567S, 2020ApJS..249...26J, 2022ApJS..259....1G}.

\citet{Sakai2008ApJ} detected high-excitation lines of the carbon-chain species, such as $cyclic$-C$_3$H$_2$, $linear$-C$_3$H$_2$, C$_4$H, C$_4$H$_2$, and CH$_3$CCH, from the low-mass protostar L1527. \citet{2010ApJ...722.1633S} found that the intensity distribution of $cyclic$-C$_3$H$_2$ shows a steep increase inward of a radius of 500 -- 1000 au from the protostar. These radii have temperatures of $\approx20-30$ K.
These carbon-chain species are not a remnant of the parent molecular cloud, but they form from CH$_4$ sublimated from dust grains around 25 K \citep{Hassel2008}.
This carbon-chain formation process was named Warm Carbon-Chain Chemistry \citep[WCCC;][]{Sakai2008ApJ}.
\citet{2017ApJ...837..174O} showed that carbon-chain species and COMs coexist around the low-mass protostar L483 but their spatial distributions are different; COMs are concentrated in the central hot corino regions, whereas carbon-chain species are extended and show a hole at the central protostar position.
Such a type of source is called a hybrid-type source.

This chemical diversity around low-mass protostars may be caused by different strengths of the interstellar radiation field (ISRF), which was proposed by \citet{2017A&A...606A..82S} based on their observations towards the starless core L1544. 
The following single-dish survey observations detected the presence of carbon-chain species and/or COMs across various low-mass protostars, including those characterized as hot corinos, WCCC sources, and hybrid-type sources.
\citet{2018MNRAS.477.4792L} found that carbon-chain-rich sources are located at the outside of the dense filaments, whereas hot-corino type sources are mainly in the dense filament where the ISRF is well shielded. 
These results are likely consistent with the scenario proposed by \citet{2017A&A...606A..82S} and interpreted as follows; the CO molecules, precursors of COMs, can survive in the dense regions and COMs become abundant, whereas CO is destroyed by the ISRF in the less shielded regions. 
The destruction of CO leads to high abundances of C and C$^+$, precursors of carbon-chain species, outside of the dense filament or edge of the molecular clouds, and these conditions are favorable for WCCC-type sources. 
Therefore, a prestellar environment could significantly modify the chemistry in the protostellar envelope, potentially promoting the formation of carbon-chain molecules with enhanced C$^+$ abundance.

Regarding high-mass protostars, studies about carbon-chain species were behind.
\citet{2014MNRAS.443.2252G} conducted survey observations of the HC$_5$N ($J=12-11$) line towards 79 high-mass protostars associated with the 6.7 GHz methanol masers. 
They detected the HC$_5$N line from 35 sources.
After these survey observations, follow-up observations were conducted.
\citet{Taniguchi2017} derived the abundances of HC$_5$N towards three high-mass protostars and found that these abundances cannot be explained by the WCCC mechanism. 
\citet{Taniguchi2019model} showed that the observed abundance ratio of HC$_5$N/CH$_3$OH around the high-mass protostar G28.28-0.36 \citep{Taniguchi2018HC5N/CH3OH} can be reproduced in their hot core model when the temperature reaches 100 K.
More recently, \citet{Taniguchi2023ApJS} presented spatial distributions of carbon-chain species (HC$_3$N, HC$_5$N, and CCH) and COMs towards five high-mass protostars obtained with the Atacama Large Millimeter/submillimeter Array (ALMA) Band 3, and indicated that HC$_5$N exists in the hot-core regions where the temperature is above 100 K. 
Based on these findings, they proposed Hot Carbon-Chain Chemistry (HCCC) to explain the observational results around high-mass protostars. 
In the HCCC mechanism, carbon-chain species are formed in the warm gas, adsorbed onto dust grains and accumulated in ice mantles below 100 K, and these carbon-chain species evaporate into the gas phase when the temperature reaches 100 K.
Stable carbon-chain species such as cyanopolyynes (HC$_{2n+1}$N, $n=1,2,3,...$) are relatively abundant than unstable radical-type carbon-chain species ($e.g,$ CCH, CCS) in HCCC, compared to the WCCC \citep[for a review see][]{2024Ap&SS.369...34T}.

Although astrochemical studies towards low-mass and high-mass protostars have been growing, our knowledge about the chemical compositions around intermediate-mass (IM) protostars ($2 M_{\odot}<m_*<8 M_{\odot}$) remains limited.
\citet{2010A&A...518A..52A} investigated the CO depletion and N$_2$H$^+$ deuteration towards Class 0 IM protostars with the IRAM 30 m telescope.
They could fit the C$^{18}$O ($J=1-0$) maps assuming that the C$^{18}$O abundance decreases inwards within the protostellar envelope until the temperatures of gas and dust reach $\approx20-25$ K, corresponding to the sublimation temperature of CO.
The deuterium fractionation of N$_2$H$^+$ was found to be 0.005--0.014, which are lower than those in prestellar clumps by a factor of 10.
The chemical compositions of COMs have been investigated towards only a few IM protostars.
\citet{2014A&A...568A..65F} observed the IM protostar NGC\,7129 FIRS 2 with the IRAM Plateau de Bure Interferometer (PdBI) and IRAM 30 m telescope, and detected numerous COMs ($e.g.,$ CH$_3$OCHO, CH$_3$CH$_2$OH, CH$_2$OHCHO, aGg'-(CH$_2$OH)$_2$, CH$_3$CH$_2$CN) from its central hot region.
They found similarities in the chemical compositions of this IM protostar with the Orion KL hot core, suggesting that the IM protostar NGC\,7129 FIRS 2 contains a hot core.
Lines of COMs have been detected from another IM protostar Cepheus E \citep{2018A&A...618A.145O}.
\citet{2018A&A...618A.145O} observed this source with the IRAM 30 m telescope and NOEMA interferometer and detected various COMs including large species such as CH$_3$COCH$_3$ and C$_2$H$_5$CN. 

Although it has been shown that hot corino chemistry emerges around IM protostars, it is still unclear whether carbon-chain molecules are formed in warm and/or hot regions ($i.e.,$ WCCC and HCCC proceed) and whether chemical diversity emerges as well as low-mass and high-mass regimes.
To address the current open questions, we need observations of carbon-chain species around IM protostars and an investigation of their relative abundances to COMs. 

This paper presents the Q-band (31.5--50 GHz) line survey observations towards 11 mainly intermediate-mass protostars 
with the Yebes 40 m telescope.
We particularly focus on carbon-chain molecules, whose rotational transition lines can be efficiently observed in the Q band.
We aim to reveal whether carbon-chain molecules are formed in warm gas around IM protostars.
Modelling of the structure of IM protostellar envelopes by \citet{2010A&A...516A.102C} shows that the radius of the 30-K dust and gas region is approximately 0.01 -- 0.02 pc.
The paper is organized as follows.
Sect. \ref{sec:obs} explains details of the observations with the Yebes 40 m telescope.
The results and spectral analyses are presented in Sects. \ref{sec:res} and \ref{sec:ana}, respectively.
We discuss carbon-chain chemistry and the chemical characteristics around IM protostars by comparing them with low-mass and high-mass regimes in Sect. \ref{sec:4}.
Our main conclusions are summarized in Sect. \ref{sec:con}.

\section{Observations} \label{sec:obs}

We have carried out Q-band (31.5--50 GHz) line survey observations with the Yebes 40 m radio telescope (Proposal IDs: 22A008 and 22B005, PI: Kotomi Taniguchi).
Eleven target protostars were selected from a sub-sample of the source list of the SOFIA Massive (SOMA) Star Formation project \citep{2017ApJ...843...33D, 2020ApJ...904...75L} with the following criteria; (1) the source declination is above $+20\degr$, and (2) other infrared (IR) sources are not contaminated within the Yebes beam size ($\approx 40\arcsec - 50\arcsec$).

Table~\ref{table:sourcelist} summarizes details of target sources. 
The coordinates correspond to the beam center of our observations.
We list protostellar properties derived by \citet{2023ApJ...942....7F} from spectral energy distribution (SED) fitting.
Note that $L_{\rm bol}$ is the intrinsic bolometric luminosity of the source, which can be different from the luminosity inferred from the received bolometric flux and assuming isotropic emission ($L_{\rm bol,iso}$) that is often quoted in observational studies of the protostars. This is because the received bolometric flux is affected by the orientation of the protostar, i.e., the ``flashlight effect'', and by foreground extinction.

The sub-sample mainly consists of IM protostars, and central values of available stellar masses ($m_*$) are within the IM regime (Table \ref{table:sourcelist}).
However, we note that IRAS\,23385+6053 has been categorized previously as a high-mass protostar \citep{2023A&A...673A.121B}.
In the end, our target source list consists of 10 IM protostars and one high-mass source (IRAS\,23385+6053).
We abbreviate IRAS source names as I and the first five numbers before ``+'' in the following parts of this paper (e.g., I\,00420).
Five sources (Cepheus E, L1206, HH288, I\,00420, and I\,20434,) were observed in the 22A008 program, and the other seven sources were observed in the 22B005 program.
The observations were carried out on February 5--14 2022 (22A008), and between September 2022 and January 2023 (22B005).

We employed the standard position-switching mode. 
The off-source positions were regions where the visual extinction ($A_V$) is below 3 mag in the $A_v$ maps obtained from the Atlas and Catalogue of Dark Clouds \citep{2005PASJ...57S...1D}\footnote{\url{https://darkclouds.u-gakugei.ac.jp/more/readme.html}}. 

\begin{table*}
\caption{Summary of 11 Target Sources}             
\label{table:sourcelist}     
\centering                        
\begin{tabular}{l l l c c c c c c c}        
\hline\hline                
Source name & R.A. (J2000) & Decl. (J2000) & $V_{\rm {LSR}}$ & {$L_{\rm {bol}}$} & $M_{\rm {env}}$ & $m_*$ & $d$ & Class \\   
            &              &               &  (km\,s$^{-1}$)  & ($L_{\sun}$)    & ($M_{\sun}$) & ($M_{\sun}$)   & (kpc) & \\
\hline                       
Cepheus E               & 23:03:13.6 & +61:42:43.5 & -11 & $6.6^{+6.7}_{-3.3}\times10^2$ & $2.2^{+2.2}_{-1.1}$ & $3.0_{-0.9}^{+1.3}$ & 0.73$^{(n)}$ & 0$^{(f)}$ \\
L1206                   & 22:28:51.4 & +64:13:41.1 & -11$^{\rm{(a)}}$ & $4.3^{+4.0}_{-2.1}\times10^3$ & $13^{+15}_{-7.1}$ & $3.4^{+3.1}_{-1.6}$ & 0.8 & 0/I$^{(g)}$ \\
HH288                   & 00:37:13.6 & +64:04:15.0 & -29 & $1.1^{+1.7}_{-0.7}\times10^3$ & $6.9^{+13}_{-4.6}$ & $3.1^{+2.6}_{-1.4}$ & 2.0 & 0$^{(h)}$ \\
IRAS\,00420+5530        & 00:44:58.0 & +55:47:00.0 & -51 & $1.5^{+3.5}_{-1.0}\times10^3$ & $17^{+38}_{-12}$ & $3.4^{+3.4}_{-1.7}$ & 2.2 & 0/I$^{(i)}$ \\
IRAS\,20343+4129\,S1    & 20:36:07.5 & +41:40:09.1 & +11.5 & $1.7^{+1.8}_{-0.9}\times10^4$$^{\rm{(o)}}$ & $9.8^{+9.7}_{-4.9}$$^{\rm{(o)}}$ & $10.9^{+5.7}_{-3.8}$$^{\rm{(o)}}$ & 1.4 & I$^{(j)}$ \\
IRAS\,00259+5625        & 00:28:42.0 & +56:42:00.0 & ...$^{\rm{(b)}}$ & $2.5^{+12.5}_{-2.0}\times10^3$ & $25^{+65}_{-18}$ & $3.3^{+6.0}_{-2.1}$ & 2.5 & 0$^{(i)}$ \\
IRAS\,05380+2020        & 05:40:54.0 & +20:22:45.0 & ...$^{\rm{(b)}}$ & $7.94\times10$$^{(d)}$ & ... & 3.3--3.5$^{(d)}$ & 1.34 & 0/I$^{(k)}$ \\
IRAS\,20293+3952        & 20:31:10.7 & +40:03:10.7 & +6.3 & $3.1^{+7.1}_{-2.2}\times10^4$$^{(p)}$ & $18.6^{+37.6}_{-12.5}$$^{(p)}$ & $13.7^{+11.6}_{-6.3}$$^{(p)}$ & 1.4 & ... &\\
IRAS\,21307+5049        & 21:32:30.6 & +51:02:16.5 & -46.6 & $9\times10^3$$^{(c)}$ & ...& ... &  5.2 & ... \\
IRAS\,22198+6336        & 22:21:26.8 & +63:51:37.6 & -11 & $1.6^{+0.7}_{-0.5}\times10^3$ & $1.9^{+1.1}_{-0.7}$ & $3.6^{+1.1}_{-0.8}$ &  0.76 & 0$^{(l)}$ \\
IRAS\,23385+6053$^{\rm{(e)}}$        & 23:40:54.5 & +61:10:28.1 & -51 & $5.6^{+27.4}_{-4.7}\times10^3$ & $26^{+52}_{-17}$ & $5.1^{+9.6}_{-3.3}$ & 4.9 & 0$^{(m)}$ \\
\hline                                   
\end{tabular}
\tablefoot{Bolometric luminosity ($L_{\rm {bol}}$), envelope mass ($M_{\rm {env}}$), stellar mass ($m_*$) are taken from \citet{2023ApJ...942....7F}. (a) The velocity was derived by the 6.7 GHz methanol maser line \citep{2009A&A...507.1117X}. On the other hand, the CO($J=1-0$) spectrum shows a peak around -10 km\,s$^{-1}$ \citep{1989ApJ...342L..87S}. (b) No available data for the systemic velocity. 
(c) Taken from the RMS Database Server\footnote{\url{http://rms.leeds.ac.uk/cgi-bin/public/RMS_DATABASE.cgi}}. 
(d) Taken from \citet{2014ApJ...784..111L}. (e) This source has been categorized as a high-mass protostar \citep{2023A&A...673A.121B}. The dynamical mass was derived to be $\sim9$ M$_{\sun}$ by fitting a Keplerian rotation disk seen in the CH$_3$CN lines \citep{2019A&A...627A..68C}. (f) Taken from \citet{2022A&A...662A.104D}. They considered its luminosity at 100 $L_{\sun}$. (g) This source was suggested to be between Class 0 and I by \citet{2020ApJ...904...75L}, whereas \citet{2023ApJ...944..135F} proposed that this is Class I. (h) Taken from \citet{2001A&A...375.1018G}. (i) Taken from \citet{2023ApJ...942....7F}. (j) Taken from \citet{2007A&A...474..911P}. (k) Taken from \citet{2014ApJ...784..111L}. (l) Taken from \citet{2010ApJ...721L.107S}. (m) Taken from \citet{1998ApJ...505L..39M}. (n) An alternative distance of 820 pc has been used for Cep E by \citet{2022A&A...662A.104D}, which is based on the measurement of the distance of the Cep OB3b cluster at $819\pm16$ pc by \citet{2019ApJ...871...46K}. Here we have retained the distance that was used in the SOMA SED fitting analysis, but acknowledge that, as is typical for most star-forming regions, distances can be uncertain by at least $\sim10\%$.  (o) Taken from the SOMA V paper by Telkamp et al. (in prep.). (p) Obtained by the SED fitting with the same method developed by \citet{2023ApJ...942....7F}.
}
\end{table*}

The Q-band receiver, one of the Nanocosmos receivers \citep{2021A&A...645A..37T}, was used for the observations.
This receiver obtains dual-polarization (H and V) data.
The Fast Fourier transform spectrometers with the 38 kHz resolution and 2.5 GHz bandwidth mode were used.
Eight base bands for each polarization were allocated, and the 31.5--50 GHz band was observed simultaneously.
The frequency resolution of 38 kHz corresponds to $\sim0.3$ km\,s$^{-1}$ in the Q band.
The main beam efficiencies ($\eta_{\rm {MB}}$) and beam sizes (Half Power Beam Width; HPBW) were approximately 50 -- 65 \% and 36\arcsec -- 54\arcsec\, between 32 GHz and 49 GHz, respectively.
The calibration was performed at the beginning of the position-switching observing the sky and both hot and cold loads, and repeating this procedure every 18 min. 
Pointing and focus were corrected every hour through pseudo-continuum observations of intense SiO maser lines towards evolved stars close to our target sources.
The pointing errors were within 7\arcsec and the calibration uncertainties are estimated to be less than 15\%.

The obtained antenna temperature ($T_{\rm{A}}^*$) is converted to the main beam temperature ($T_{\rm{MB}}$) by using the following formulae \citep{2021A&A...645A..37T}; $T_{\rm{MB}} = T_{\rm{A}}^*\frac{\eta_{\rm {F}}}{\eta_{\rm {MB}}}$, where $\eta_{\rm {F}}$ is the forward efficiency \citep[0.91--0.93 at the Q-band;][]{2021A&A...645A..37T}.

\section{Results and analyses} \label{sec:3}

\subsection{Results} \label{sec:res}
We made fits files of the spectra from CLASS (software of the GILDAS package), and further data reduction was conducted with the CASSIS software \citep{CASSIS}. 
Figs. \ref{fig:speccarbon3}--\ref{fig:specS} in Appendix \ref{sec:append1} show spectra of carbon-chain species (HC$_3$N, HC$_5$N, C$_3$H, $linear$ ($l$)-H$_2$CCC, $cyclic$ ($c$)-C$_3$H$_2$, C$_4$H, CCS, C$_3$S, and CH$_3$CCH), COMs (CH$_3$OH, CH$_3$CHO, and CH$_3$CN), H$_2$CCO, HNCO, and sulfur (S)-bearing species ($^{13}$CS, C$^{34}$S, HCS$^+$, and H$_2$CS) towards the 11 sources.
We categorized CCS and C$_3$S into carbon-chain species by following the definitions of carbon-chain species \citep{2024Ap&SS.369...34T}, even though they contain a sulfur atom.
Table \ref{table:linesum} in Appendix \ref{sec:append1} summarizes information on each line (transition, rest frequency, and upper-state energy).
The average rms noise levels measured in line-free channels are around 5 mK.

\begin{table*}[t]
\caption{Summary of detection status in IM protostars}
\label{table:quick}
\begin{center}
\small
\begin{tabular}{l c c c c c c c c c c c}
\hline
\hline
Species & Cepheus E & L1206 & HH288 & I\,00420 & I\,20343 & I\,00259 & I\,05380 & I\,20293 & I\,21307 & I\,22198 & I\,23385  \\
\hline
\multicolumn{12}{l}{\bf {Carbon-Chain Molecules}} \\
HC$_3$N          & $\surd$ & $\surd$ & $\surd$ & $\surd$ & $\surd$ & $\surd$ & $\surd$ & $\surd$ & $\surd$ & $\surd$ & $\surd$ \\
HC$_5$N          & $\surd$ & $\surd$ & $\surd$ & $\surd$ & $\surd$ & $\surd$ &         & $\surd$ &         & $\surd$ & $\surd$ \\
C$_3$H           &         & $\surd$ & $\surd$ & $\surd$ &         & $\surd$ &         &         &         & $\surd$ &         \\
C$_4$H           & $\surd$ & $\surd$ & $\surd$ & $\surd$ & $\surd$ & $\surd$ & $\surd$ &         & $\surd$ & $\surd$ & $\surd$ \\
$l$-H$_2$CCC     & ($\surd$) & $\surd$ & $\surd$ & $\surd$ & $\surd$ & ($\surd$) &         &         &         & $\surd$ & $\surd$ \\
$c$-C$_3$H$_2$ & $\surd$ & $\surd$ & $\surd$ & $\surd$ & $\surd$ & $\surd$ &         & $\surd$ &         & $\surd$ & $\surd$ \\
CCS                & $\surd$ & $\surd$ & $\surd$ & $\surd$ & $\surd$ & $\surd$ & $\surd$ & $\surd$ &         & $\surd$ & $\surd$ \\
C$_3$S           & $\surd$ & $\surd$ & $\surd$ &         &         &         &         &         &         & $\surd$ &         \\
CH$_3$CCH        &         & $\surd$ & $\surd$ &         & $\surd$ &         &         & $\surd$ &         & $\surd$ &         \\
\hline
\multicolumn{12}{l}{\bf {COMs}} \\
CH$_3$OH         & $\surd$ & $\surd$ & $\surd$ & $\surd$ & $\surd$ & $\surd$ &         & $\surd$ & $\surd$ & $\surd$ & $\surd$ \\
CH$_3$CHO        & $\surd$ & $\surd$ & $\surd$ & $\surd$ & $\surd$ & $\surd$ &         & $\surd$ &         & $\surd$ & $\surd$ \\
CH$_3$CN         & $\surd$ & $\surd$ & $\surd$ & $\surd$ & $\surd$ & $\surd$ &         &         &         & $\surd$ &         \\
\hline
H$_2$CCO         & $\surd$ & $\surd$ & $\surd$ & $\surd$ & $\surd$ & $\surd$ &         &         &         & $\surd$ &         \\
HNCO               & $\surd$ & $\surd$ & $\surd$ & $\surd$ & $\surd$ & $\surd$ &         & $\surd$ &         & $\surd$ &         \\
\hline 
\multicolumn{12}{l}{\bf {S-bearing species}} \\
$^{13}$CS          & $\surd$ & $\surd$ & $\surd$ & $\surd$ & $\surd$ & $\surd$ &         & $\surd$ & $\surd$ & $\surd$ & $\surd$ \\
C$^{34}$S          & $\surd$ & $\surd$ & $\surd$ & $\surd$ & $\surd$ & $\surd$ &         & $\surd$ & $\surd$ & $\surd$ & $\surd$ \\
HCS$^+$          & $\surd$ & $\surd$ & $\surd$ & $\surd$ & $\surd$ & $\surd$ &         & $\surd$ & $\surd$ & $\surd$ & $\surd$ \\
H$_2$CS          & $\surd$ & $\surd$ & $\surd$ & $\surd$ & $\surd$ & $\surd$ &         & $\surd$ & $\surd$ & $\surd$ & $\surd$ \\
\hline
\end{tabular}
\end{center}
\tablefoot{$l$-H$_2$CCC has been tentatively detected with S/N ratio of 3 in Cepheus E and I\,00259 and then we indicate as ($\surd$).}
\end{table*}

Table \ref{table:quick} summarizes the detection status in each source.
Cyanoacetylene (HC$_3$N) has been detected from all of the sources, and cyanodiacetylene (HC$_5$N) is detected in almost all of the sources, except for I\,05380 and I\,21307.
$c$-C$_3$H$_2$ is associated with the protostars where HC$_5$N has been detected.
Two $c$-C$_3$H$_2$ lines show different features, which are likely caused by different upper-state energies (Table \ref{table:linesum}).
CCS is detected from all of the sources except for I\,21307.
All of the carbon-chain species listed in Table \ref{table:quick} have been detected from L1206, HH288, and I\,22198.
The other sources, except for I\,05380 and I\,21307, show lines from at least five carbon-chain species. 
Only three and two carbon-chain species have been detected from I\,05380 and I\,21307.

In addition to carbon-chain species, three COMs (CH$_3$OH, CH$_3$CHO, and CH$_3$CN), H$_2$CCO, and HNCO have been detected in the Q band.
Methanol (CH$_3$OH), one of the most fundamental COMs, and S-bearing species are detected from all of the sources except for I\,05380.
We can see the wing emission in the spectra of CH$_3$OH in Cepheus E, I\,00259, and I\,20293 (Figs.\ref{fig:specCH3OH1}--\ref{fig:specCH3OH3}).
Except for the above three sources, the $4_{1,4}-3_{0,3}$ $E$ lines of CH$_3$OH show a single peak and the line peak coincides with the rest frequency, which means that the emission comes from low-velocity quiescent gas, presumably envelopes.

The IM protostars L1206, HH288, and I\,22198 are the most line-rich sources, whereas I\,05380 is likely a line-poor source.
The source distance of I\,05380 is 1.34 kpc (Table \ref{table:sourcelist}), and this source is not the farthest one, which means that the beam dilution effect (a beam size of $\approx 40\arcsec$ corresponds to 0.25 pc at 1.34 kpc) does not cause the non-detection of molecular lines.
This source has the lowest luminosity among our target IM protostars, and the gas and dust temperatures could be lower.
Thus, hot/warm regions are smaller compared to the other sources.
These physical conditions may cause a limited number of the detected species.

\subsection{Spectral analyses} \label{sec:ana}

We derived rotational temperatures using seven HC$_5$N lines (from $J=12-11$ to $J=18-17$) and four CH$_3$OH lines in the Q band (Sect. \ref{sec:rot}).
The rotational temperature provides a hint of where carbon-chain species exist; outer cold envelopes, lukewarm envelopes, or central hot core regions.
Such a distinguishment is important to constrain the formation processes of carbon-chain species around IM protostars; just a remnant of the parent molecular cloud, WCCC, or HCCC \citep{2024Ap&SS.369...34T}.

We analyzed spectra and derived column densities of the other species with the Markov Chain Monte Carlo (MCMC) method assuming the Local Thermodynamic Equilibrium (LTE) condition because there is not enough data to conduct the rotational diagram analysis (Sect. \ref{sec:MCMC}).

\subsubsection{Rotational diagram of HC$_5$N and CH$_3$OH} \label{sec:rot}

\begin{figure*}[ht]
   \centering
    \includegraphics[scale = 0.8]{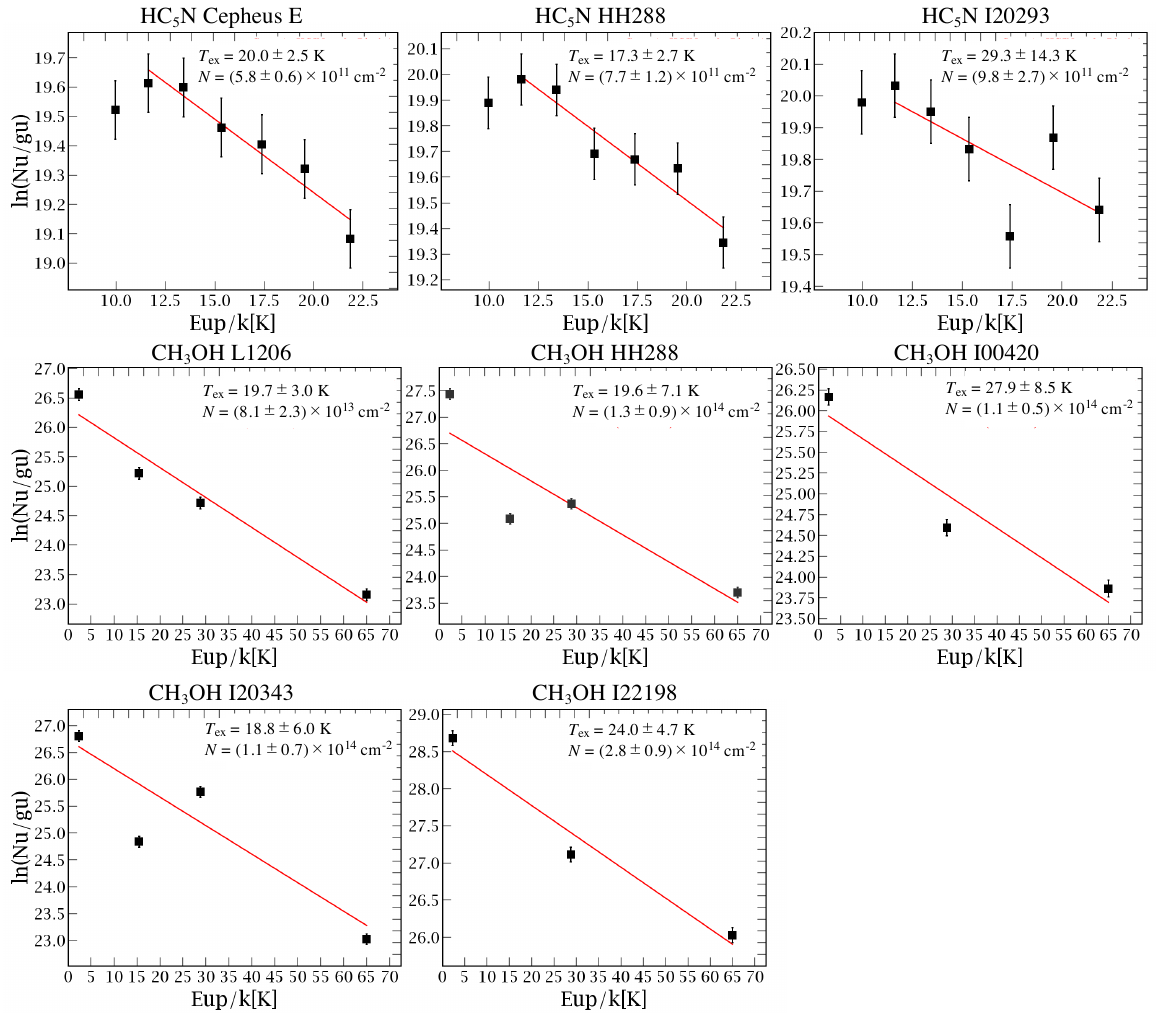}
   \caption{Rotational diagrams of HC$_5$N and CH$_3$OH. Errors for each data point indicate 10\% errors.} \label{fig:RD}
\end{figure*}

We fitted the spectra with a Gaussian profile and conducted the rotational diagram analysis in the CASSIS software \citep{CASSIS}.
We applied this method for all of the sources where the HC$_5$N lines have been detected.
However, we can fit the data and derive rotational temperatures only in three sources, Cepheus E, HH288, and I\,20293.
We could not derive the rotational temperatures in the other sources because the data points cannot be fitted by this method.
This is likely caused by low signal-to-noise (S/N) ratios of the lines and/or non-Gaussian profiles.

The top panels of Fig. \ref{fig:RD} show the rotational diagrams of HC$_5$N in these three sources.
The $J=12-11$ line shows systematically lower values in all of the sources and could not be fitted with the other lines simultaneously.
This is caused by the fact that the lowest $J$ line was observed at the edge of the band of the receiver, and some systematic effects cause the intensity fluctuation.
It is likely that HC$_5$N is located in both cold envelopes and warm envelopes; contributions from cold envelopes are larger for the lowest energy line.
Since its spatial distributions are unknown, we cannot estimate the beam dilution effect for each line.
Thus, we could not correct the beam-filling factors, and then we derived the average rotational temperatures within the beams. 
We need observations of lower $J$ lines to cover cold-envelope components.

We fitted the data excluding the $J=12-11$ line to avoid the systematic effects that we mentioned in the preceding paragraph.
The rotational temperatures were derived to be $20.0\pm2.5$ K, $17.3\pm2.7$ K, and $29.3\pm14.3$ K in Cepheus E, HH288, and I\,20293, respectively.
The rotational temperatures in Cepheus E and HH288 are well constrained, and we applied these temperatures in the analyses of the other carbon-chain species (Sect. \ref{sec:MCMC}).

We conducted rotational diagram analysis for the CH$_3$OH data.
The spectra in several sources show non-Gaussian profiles, such as wing emission or complicated several velocity components, and we could not obtain rotational temperatures in these cases.
We could conduct this method towards five sources that show Gaussian profiles.
The middle and bottom panels of Fig. \ref{fig:RD} show the rotational diagrams of these five sources.
The derived rotational temperatures of CH$_3$OH are around 20--30 K, which are similar to those of HC$_5$N.
Since the observed lines of CH$_3$OH have low upper-state energies (Table \ref{table:linesum}) and the line widths are not quite wide, the emission likely comes from low-velocity quiescent gas in the envelope \citep{Taniguchi2020MNRAS, 2021A&A...655A..65T,2024ApJ...960..127G}.

\subsubsection{Markov Chain Monte Carlo (MCMC) method} \label{sec:MCMC}

We conducted the MCMC spectral analysis in the CASSIS software \citep{CASSIS}.
We assumed the LTE condition for all of the species. 
In the fitting procedure, we treated the molecular column density ($N$), line width (FWHM), and centroid velocity ($V_{\rm {LSR}}$) as free parameters.
Since we do not know the molecular spatial distributions, we derived the beam-averaged column densities.

In analyses of carbon-chain species except for HC$_5$N and CH$_3$CCH, we fixed the excitation temperatures ($T_{\rm {ex}}$), because we could not determine both column density and excitation temperature simultaneously due to insufficient lines.
The excitation temperatures of 20 K are applied for all of the sources except for HH288, in which we applied the rotational temperature of HC$_5$N (17.3 K; Sect. \ref{sec:rot}). 
We could not fit all of the lines of CCS, $c$-C$_3$H$_2$, and $l$-H$_2$CCC simultaneously under an assumption with the single excitation temperature of 20 K.
In that case, we fitted the spectra dividing into two groups depending on the upper state energies; lines with low upper state energies ($E_{\rm {up}}/k \leq 9$ K) are fitted with the excitation temperature of 10 K, whereas those with high upper state energies ($E_{\rm {up}}/k \geq 12$ K) are fitted with the excitation temperature of 20 K.
We indicated these different assumed excitation temperatures as (low) and (high), respectively, in Table \ref{table:column}.
The assumed excitation temperature of 10 K is a typical gas kinetic temperature of starless cores.
Such a method was applied because we assumed that carbon-chain molecules are likely present in both outer cold envelopes and warm envelopes which are close to the IM protostars.

The $l$-H$_2$C$_3$ has been detected tentatively in Cepheus E, and then we treated its column density as the upper limit.
We fitted four C$_4$H lines simultaneously because they have similar upper-state energies. We did not fit lines with non-Gaussian profiles or low S/N ratios. However, all of the lines cannot be well fitted simultaneously: 
the fit of the $N=5-4$ line emission fails to reproduce the $N=4-3$ line emission, and vice versa.
Only the best-fitting results which show the smallest chi-square values are displayed in the spectral figures. 
However, the derived physical parameters are calculated considering this problem; large errors are included in the derived physical parameters if all of the lines have not been fitted simultaneously.

In the case of HC$_5$N, we treated the excitation temperature as an additional free parameter because its seven lines are available, which means that its column densities and excitation temperatures were determined simultaneously.
In I\,00259 and I\,23385, the S/N ratios are not high enough, or non-detection of several lines, and then we fixed the excitation temperature at 20 K.
We could not fit the $J=12-11$ line with the other lines simultaneously due to systematically low intensities (see also Sect. \ref{sec:rot}).
Then, we fitted the $J=12-11$ transition with a fixed excitation temperature of 10 K, but they are reference values due to the uncertainties in peak intensities.
In Table \ref{table:column}, these column densities are labeled as (low).
The other lines are fitted with the excitation temperature as a free parameter.
The determined excitation temperatures and column densities are listed in Tables \ref{table:Tex} and \ref{table:column}, respectively.

We derived the column densities and excitation temperatures of CH$_3$CCH using two $K$-ladder lines ($K=0$ and 1) with the MCMC method in the five sources.
The rotational temperatures are derived to be around 20 K, which are consistent with those of HC$_5$N (Table \ref{table:Tex}).
The results also support that carbon-chain species exist in warm regions, because CH$_3$CCH is suggested to be enhanced by the WCCC mechanism \citep{Taniguchi2019model}.

In analyses of COMs and S-bearing species, we applied the excitation temperature of 20 K, which is constrained by the rotational diagram analyses of CH$_3$OH (Sect. \ref{sec:rot}).
We treated the excitation temperature as a free parameter of CH$_3$OH data in I\,21307 and I\,22198, in which two lines with the Gaussian profile ($4_{1,4}-3{-0,3}$ $E$ and $1_{0,1}-0_{0,0}$ $A$) have been detected. 
The derived excitation temperatures are summarized in Table \ref{table:Tex}.
In I\,23385, the excitation temperature was fixed at 20 K.
We excluded spectra with low S/N ratios and non-Gaussian profiles from the fitting.
We analyzed spectra with the two velocity components for CH$_3$CN in I\,20293, and $^{13}$CS, C$^{34}$S, and H$_2$CS in I\,23385. 
The two velocity components in I\,23385 are consistent with those found in the C$^{18}$O and C$^{17}$O lines \citep[-50.5 km\,s$^{-1}$ and -47.8 km\,s$^{-1}$;][]{2004A&A...414..299F}.
Although CH$_3$CN has two $K$-ladder lines, the $K=1$ line was detected with low S/N ratios and we could not use it for fitting. 
We thus fitted the $K=0$ line with a fixed excitation temperature in the CH$_3$CN analysis.

The derived column densities are summarized in Table \ref{table:column}.
Some column densities show large uncertainties due to low S/N ratios or non-Gaussian line features.
The derived line widths and centroid velocities are summarized in Table \ref{table:vlsr} in Appendix \ref{sec:append1}.

\begin{sidewaystable*}
\caption{Summary of column densities.}\label{table:column}
\scalebox{0.79}{
\begin{tabular}{lccccccccccc} 
\hline\hline             
Species & Cepheus E & L1206 & HH288 & I\,00420 & I\,20343 & I\,00259 & I\,05380 & I\,20293 & I\,21307 & I\,22198 & I\,23385 \\
 & $N$ (cm$^{-2}$) & $N$ (cm$^{-2}$) & $N$ (cm$^{-2}$) & $N$ (cm$^{-2}$) & $N$ (cm$^{-2}$) & $N$ (cm$^{-2}$) & $N$ (cm$^{-2}$) & $N$ (cm$^{-2}$) & $N$ (cm$^{-2}$) & $N$ (cm$^{-2}$) & $N$ (cm$^{-2}$) \\ 
\hline
HC$_3$N & $1.9 (0.3)\times10^{13}$ & $2.9(0.4)\times10^{13}$ & $3.7(0.3)\times10^{13}$ & $7.2(2.7)\times10^{12}$ & 
$4.2(0.4)\times10^{13}$ & $1.2(0.3)\times10^{13}$	& $1.6(1.0)\times10^{12}$ & $4.8(1.0)\times10^{13}$ & $6.0(3.1)\times10^{12}$ & $3.6(0.4)\times10^{13}$ & $2.5(0.4)\times10^{13}$ \\
HC$_5$N (low) & $8.3(4.4)\times10^{12}$ & $9.6(5.8)\times10^{12}$ & $8.1(3.6)\times10^{12}$ & ... & $7.7(3.7)\times10^{12}$	& ... & ... & $7.0(3.3)\times10^{12}$ & ... & $9.3(5.2)\times10^{12}$ & $4.5(2.7)\times10^{12}$ \\
HC$_5$N & $4.5(2.4)\times10^{12}$ & $6.0(2.6)\times10^{12}$ & $4.4(2.5)\times10^{12}$ & $7.9(5.8)\times10^{11}$ & $5.3(2.9)\times10^{12}$ & $1.2(0.8)\times10^{12}$ & ... & $3.6(2.1)\times10^{12}$ & ... & $7.8(3.0)\times10^{12}$ & $2.6(1.6)\times10^{12}$ \\
C$_4$H & $1.3(0.7)\times10^{13}$ & $3.4(0.9)\times10^{13}$ & $2.3(0.7)\times10^{13}$ & $1.2(0.7)\times10^{13}$ & $1.2(0.8)\times10^{13}$ & $2.2(1.2)\times10^{13}$ & $9.4(6.3)\times10^{12}$ & ... & $2.2(1.4)\times10^{13}$ & $3.2(1.7)\times10^{13}$ & $6.0(2.4)\times10^{13}$ \\
C$_3$H & ... & $4.1(2.5)\times10^{13}$ & $1.7(1.0)\times10^{13}$ & $1.6(1.0)\times10^{13}$ & ... & $1.7(1.1)\times10^{13}$ & ... & ... & ...& $1.6(0.9)\times10^{13}$ & ...\\
CCS (high) & $1.7(1.1)\times10^{13}$ & $1.4(0.9)\times10^{13}$ & $1.3(0.8)\times10^{13}$ & $7.4(4.7)\times10^{12}$ & ... & ... & ... & ... & ... & $1.1(0.7)\times10^{13}$ & ... \\
CCS (low) & $1.8(0.5)\times10^{13}$ & $1.3(0.5)\times10^{13}$ & $2.6(0.6)\times10^{13}$ & $3.8(2.2)\times10^{12}$ & $8.2(4.8)\times10^{12}$ & $7.4(4.0)\times10^{12}$ & $2.5(1.4)\times10^{12}$ & $6.7(4.2)\times10^{12}$ & ... & $1.2(0.4)\times10^{13}$ & $7.0(0.4)\times10^{12}$ \\
& & & & & & & & & & & $7.2(0.2)\times10^{12}$$^{(b)}$ \\
C$_3$S & $3.0(1.8)\times10^{12}$ & $1.9(1.2)\times10^{12}$ & $2.5(1.5)\times10^{12}$ & ... & ... & ... & ... & ... & ... & $2.4(1.5)\times10^{12}$ & ... \\
$c$-C$_3$H$_2$ (high) & $1.1(0.7)\times10^{13}$ & $2.5(1.4)\times10^{13}$ & $1.4(0.9)\times10^{13}$ & ... & $1.7(1.0)\times10^{13}$ & $6.8(4
5)\times10^{12}$ & ... & $1.3(0.8)\times10^{13}$ & ... & $2.1(1.3)\times10^{13}$ & ... \\
$c$-C$_3$H$_2$ (low) & $3.4(1.8)\times10^{13}$ & $6.9(2.7)\times10^{13}$ & $5.0(2.4)\times10^{13}$ & $1.8(1.1)\times10^{13}$ & $4.0(2.3)\times10^{13}$ & $1.9(1.2)\times10^{13}$ & ... & $3.3(1.8)\times10^{13}$ & ... & $4.6(2.4)\times10^{13}$ & $3.6(2.1)\times10^{13}$ \\
$l$-H$_2$CCC (high) & ... & $8.8(5.3)\times10^{12}$ & $2.9(1.8)\times10^{12}$ & $5.2(3.4)\times10^{12}$ & $6.9(4.3)\times10^{12}$ & ... & ... & ... & ... & $6.5(3.9)\times10^{12}$ & ... \\
$l$-H$_2$CCC (low) & $<2.6\times10^{12}$ & $3.8(2.2)\times10^{12}$ & $3.7(2.2)\times10^{12}$ & $2.9(1.8)\times10^{12}$ & ... & $<3.6\times10^{12}$ & ... & ... & ... & $3.4(2.0)\times10^{12}$ & ... \\
CH$_3$CCH & ... & $2.4(1.4)\times10^{14}$ & $2.9(1.7)\times10^{14}$ & ... & $5.4(2.9)\times10^{14}$ & $1.1(0.7)\times10^{14}$ & ... & $3.6(2.1)\times10^{14}$ & ... & $2.1(1.3)\times10^{14}$ & ... \\
CH$_3$OH & & $8.1(2.3)\times10^{13}$$^{(a)}$ & $1.3(0.9)\times10^{14}$$^{(a)}$ & $1.1(0.5)\times10^{14}$$^{(a)}$ & $1.1(0.7)\times10^{14}$ & ... & ... & ... & $8.8(3.9)\times10^{14}$ & $2.5(0.8)\times10^{15}$ & $7.1(0.8)\times10^{15}$  \\
CH$_3$CHO & $6.0(3.8)\times10^{13}$ & $1.0(0.6)\times10^{14}$ & $8.0(4.3)\times10^{13}$ & $4.1(2.6)\times10^{13}$ & $6.3(3.8)\times10^{13}$ & $7.7(4.1)\times10^{13}$ & ... & $6.0(3.4)\times10^{13}$ & ... & $5.9(3.6)\times10^{13}$ & $3.9(2.4)\times10^{13}$ \\
H$_2$CCO & $5.1(3.2)\times10^{13}$ & $5.2(3.2)\times10^{13}$ & $4.9(2.9)\times10^{13}$ & $4.5(2.8)\times10^{13}$ & $5.3(3.2)\times10^{13}$ & $4.3(2.8)\times10^{13}$ & ... & ... & ... & $5.0(3.0)\times10^{13}$ & ... \\
HNCO & $1.7(1.0)\times10^{13}$ & $2.5(1.5)\times10^{13}$ & $1.8(1.1)\times10^{13}$ & $1.1(0.7)\times10^{13}$ & $1.1(0.7)\times10^{13}$ & $2.3(1.3)\times10^{13}$ & ... & $1.3(0.8)\times10^{13}$ & ... & $1.4(0.9)\times10^{13}$ & ... \\
CH$_3$CN & $5.0(3.1)\times10^{12}$ & $1.2(0.6)\times10^{13}$ & $2.9(1.7)\times10^{12}$ & $4.6(3.1)\times10^{12}$ & $7.3(4.7)\times10^{12}$ & $5.0(2.8)\times10^{12}$ & ... &  $6.2(0.2)\times10^{12}$ & $5.0(3.2)\times10^{12}$ & $6.1(3.8)\times10^{12}$ & ... \\
 &  & &  & &  & & ... & $7.9(0.3)\times10^{12}$$^{(b)}$  & \\
$^{13}$CS & $1.0(0.6)\times10^{13}$ & $1.8(0.9)\times10^{13}$ & $1.6(0.8)\times10^{13}$ & $5.8(3.9)\times10^{12}$ & $3.1(1.1)\times10^{13}$ & $6.1(3.8)\times10^{12}$ & ... & $2.1(0.9)\times10^{13}$ & $6.3(3.8)\times10^{12}$ & $1.7(0.8)\times10^{13}$ & $1.5(0.2)\times10^{13}$ \\
& & & & & & & & & & & $5.2(0.8)\times10^{12}$$^{(b)}$ \\
C$^{34}$S & $2.7(0.7)\times10^{13}$ & $5.8(0.8)\times10^{13}$ & $4.0(0.9)\times10^{13}$ & $2.1(0.7)\times10^{13}$ & $7.8(1.0)\times10^{13}$ & $2.8(0.8)\times10^{13}$ & ... & $5.7(0.9)\times10^{13}$ & $2.3(0.9)\times10^{13}$ & $6.0(0.9)\times10^{13}$ & $4.7(0.4)\times10^{13}$ \\
& & & & & & & & & & & $3.5(0.5)\times10^{13}$$^{(b)}$ \\
HCS$^+$ & $7.2(4.5)\times10^{12}$ & $1.2(0.7)\times10^{13}$ & $6.9(4.1)\times10^{12}$ & $6.1(4.0)\times10^{12}$ & $1.0(0.6)\times10^{13}$ & $5.2(3.3)\times10^{12}$ & ... & $6.9(4.4)\times10^{12}$ & $3.1(1.9)\times10^{12}$ & $1.0(0.6)\times10^{13}$ & $1.5(0.9)\times10^{13}$ \\
H$_2$CS & $1.3(0.8)\times10^{14}$ & $8.9(5.4)\times10^{13}$ & $1.6(0.9)\times10^{14}$ & $8.8(5.5)\times10^{13}$ & $1.2(0.8)\times10^{14}$ & $8.5(4.9)\times10^{13}$ & ... & $1.2(0.7)\times10^{14}$ & $7.8(4.6)\times10^{13}$ & $1.1(0.6)\times10^{14}$ & $8.8(5.4)\times10^{13}$ \\
\hline
\end{tabular}
}
\tablefoot{Numbers in parentheses indicate the standard deviation error. ``high'' and ``low'' mean the column densities derived by fixed excitation temperatures at 20 K and 6.5 K, respectively. Without these indications, we fixed the excitation temperatures at 20 K. \\(a) The column densities are derived by the rotational diagram method. \\(b) The 2nd velocity component.}
\end{sidewaystable*}

\begin{table*}
\caption{Excitation temperatures of HC$_5$N, CH$_3$OH, and CH$_3$CCH derived by the MCMC method.}             
\label{table:Tex}     
\centering                        
\begin{tabular}{lccccccccccc}        
\hline\hline                
Species & Cepheus E & L1206 & HH288 & I\,00420 & I\,20343 & I\,00259 & I\,05380 & I\,20293 & I\,21307 & I\,22198 & I\,23385 \\
\hline                       
HC$_5$N & $25 \pm8$ & $26\pm8$ & $24\pm8$ & $22\pm5$ & $27\pm9$ & 20 (fix) & ... & $24\pm8$ & ... & $25\pm8$ & 20 (fix)  \\
CH$_3$OH & ... & ... & ... & ... & ... & ... & ... & ... & $39\pm8$ & $36\pm8$ & ... \\
CH$_3$CCH &... & $16.5\pm0.8$ & $23\pm3$ &...& $19\pm4$ & ...& ...& $22\pm4$ & ... & $21\pm4$ &... \\
\hline                                   
\end{tabular}
\tablefoot{The unit is Kelvin [K]. Errors indicate the standard deviation.}
\end{table*}

\section{Discussion}\label{sec:4}

\subsection{Comparisons of rotational temperatures of HC$_5$N} \label{sec:disT}

Here we compare the rotational temperatures of HC$_5$N around IM protostars to those around low-mass and high-mass counterparts. We utilize the results obtained by single-dish telescopes and the derived rotational temperatures are beam-averaged values.

\citet{Sakai2009ApJ} conducted the Q-band observations with the Green Bank 100 m telescope (GBT) and 3 mm band (90--150 GHz) observations with the IRAM 30 m telescope towards the low-mass protostar L1527 ($d=140$ pc), which is one of the WCCC sources \citep{Sakai2008ApJ}.
The derived rotational temperature of HC$_5$N is $14.7\pm5.3$ K, using three lines ($J=16-15$, $17-16$, and $32-31$).
Note that the rotational temperature was derived by fitting with almost two data points because the upper state energies of $J=16-15$ and $J=17-16$ transitions are close.

\citet{Taniguchi2017} detected the HC$_5$N lines in the Ka-band ($J=10-9$ and $11-10$) with GBT, and in the 45 GHz ($J=16-15$ and $17-16$) and 90 GHz ($J=31-30$, $32-31$, $34-33$, $36-35$, $38-37$, and $39-38$) bands with the Nobeyama 45 m radio telescope from three high-mass protostars, or massive young stellar objects (MYSOs; G\,12.89+0.49, G\,16.86-2.16, and G\,28.28-0.36).
The rotational temperatures with the beam-size correction are $18\pm2$, $17\pm2$, and $13.8^{+1.5}_{-1.1}$ K in G\,12.89+0.49 ($d=2.94$ kpc), G\,16.86-2.16 ($d=1.7$ kpc), and G\,28.28-0.36 ($d=3.0$ kpc), respectively.
Because the observations have a low-angular resolution, these temperatures are considered to be the lower limits due to contamination from outer cold envelopes, as pointed out by \citet{Taniguchi2021}.

In the case of the IM protostars, the rotational temperatures of HC$_5$N were derived to be around 20 K (Sect. \ref{sec:rot}).
In the MCMC analysis, the derived excitation temperatures are slightly higher ($\sim25$ K) but consistent with the former within their errors.
The rotational temperatures around the IM protostars are close to those around the low-mass and high-mass protostars and clearly higher than the gas kinetic temperature in molecular clouds ($\sim 10$ K).
In addition, the excitation temperatures around the IM protostars agree with the WCCC scenario in which CH$_4$ sublimated from dust grains around 25 K forms carbon-chain species.
These results imply that carbon-chain molecules are formed in warm gas around the IM protostars by WCCC \citep{Sakai2008ApJ, Hassel2008} or HCCC \citep{Taniguchi2019model, Taniguchi2023ApJS}.
The derived excitation temperatures of CH$_3$CCH also support this scenario.

Here, we constrain which mechanism is dominant in our observations, WCCC or HCCC.
The size of the hot region with temperatures above 100 K of Cepheus E was estimated at $0.7\arcsec$, corresponding to $\sim510$ au \citep{2018A&A...618A.145O}.
Thus, our target sources should have much smaller hot regions ($T>100$ K) compared to the beam size ($40\arcsec-50\arcsec$).
This means that the detected carbon-chain emission around the IM protostars should come from warm envelopes rather than central hot regions.
Thus, it is concluded that the WCCC mechanism forms the carbon-chain species around the IM protostars.
We need high angular resolution ($\leq 0.5\arcsec$) observations to investigate whether the HCCC mechanism works around the IM protostars.

Since our observations cover lines only with upper state energies around 10 -- 22 K (Table \ref{table:linesum}), the detected emission may be biased to warm or cold components ($i.e.,$ the outer layers of the protostellar envelopes).
Even in this case, our conclusion that carbon-chain species form around the IM protostars is robust.
Since the photodissociation region (PDR) chemistry does not produce HC$_5$N efficiently, 
the detected emission of HC$_5$N likely comes from mainly warm central gas, not cavity walls of molecular outflows.

In summary, the carbon-chain formation around protostars occurs ubiquitously. 
It is difficult to conclude which formation mechanism is dominant, WCCC or HCCC, around IM protostars from the rotational temperatures derived by Q-band single-dish observations.
We need high $J$ transition observations and imaging observations with interferometers to solve the current open questions.

\subsection{Comparisons of the chemical compositions}\label{sec:dischem}

\begin{figure*}[ht]
   \centering
    \includegraphics[bb = 0 10 550 665, scale = 0.84]{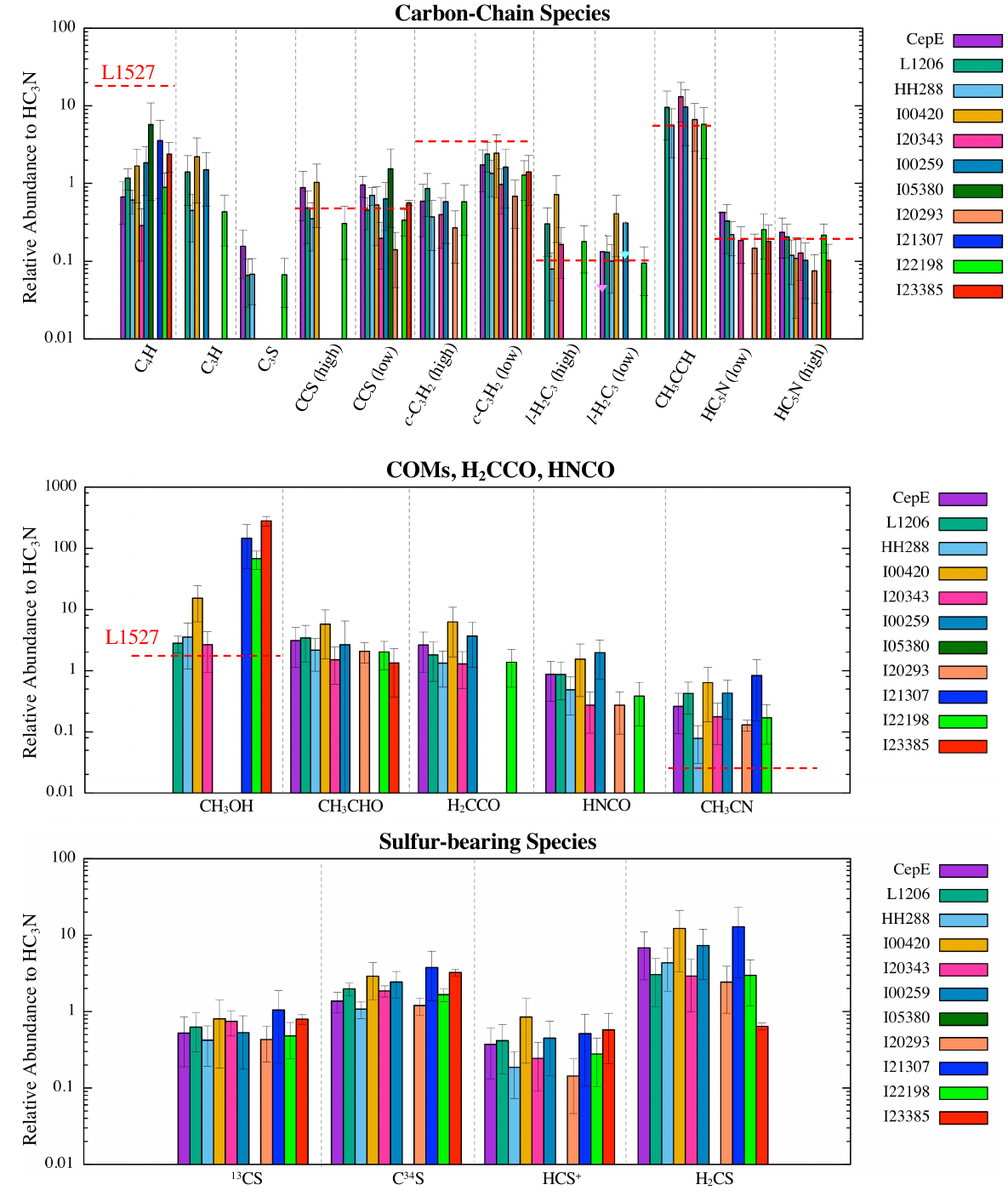}
   \caption{Comparisons of molecular abundances with respect to HC$_3$N; (top) carbon-chain species, (middle) COMs, H$_2$CCO, and HNCO, and (bottom) S-bearing species. Errors indicate the standard deviation. Note in the caption that the ``high'' and ``low'' are the high and low temperatures components. The red dashed lines indicate the abundance ratios in the low-mass WCCC source L1527 \citep{2019PASJ...71S..18Y}.} \label{fig:chemcom}
\end{figure*}

Fig. \ref{fig:chemcom} shows comparisons of molecular abundances with respect to HC$_3$N, which are defined as $N$(molecules)/$N$(HC$_3$N), among the 11 protostars.
We use HC$_3$N as the standard because it has been detected in all the sources, and it is useful when we compare to results in low-mass and high-mass regimes as we describe later.
The top, middle, and bottom panels show comparisons of carbon-chain species, three COMs, H$_2$CCO, and HNCO, and S-bearing species, respectively.
In the case that two velocity components have been derived, we plotted the sums of these two components.
For comparisons with the low-mass WCCC source L1527, we indicate the observed abundance ratios in this source \citep{2019PASJ...71S..18Y} as the horizontal red dashed lines. 

As a general trend, the derived abundances do not vary among the sources, when we focus on a particular molecule, especially the S-bearing species.
On the other hand, CH$_3$OH shows a larger variation in its abundances among three protostars; I\,21307, I\,22198, and I\,23385 show larger abundances compared to the other sources where we can derive its abundances.
As we can see wing emission in its spectra (Figs. \ref{fig:specCH3OH1}--\ref{fig:specCH3OH3} in Appendix \ref{sec:append1}), the CH$_3$OH lines come from not only warm envelopes but also molecular outflows and shock regions, as we have already mentioned in Sect. \ref{sec:rot}.

We compared the abundances with respect to HC$_3$N between the 11 protostars and the low-mass WCCC source L1527 \citep{2019PASJ...71S..18Y}. 
The C$_4$H/HC$_3$N ratio in L1527 was derived to be $\sim19$, which is higher than our target sources.
The high-temperature components of HC$_5$N in Cepheus E, L1206, and I\,22198 show similar values with that in L1527 ($\sim0.2$), whereas HH288, I\,00420, I\,20343, I\,00259, I\,20293, and I\,23385 show slightly lower values compared to L1527.
The abundances of $c$-C$_3$H$_2$ around the IM protostars tend to be lower than that in L1527 ($\approx3.6$).
The abundances of the other carbon-chain species in IM protostars are consistent with those in L1527 within errors; the abundance ratios in L1527 are CCS/HC$_3$N$\approx0.49$, $l$-H$_2$CCC/HC$_3$N$\approx0.1$, CH$_3$CCH/HC$_3$N$\approx5.7$, respectively. 
These results suggest that the formation of large carbon-chain species has not proceeded yet around most of the target IM protostars, because the WCCC mechanism starts with CH$_4$ and small carbon-chain species form first.

The CH$_3$OH/HC$_3$N ratio in L1527 is around 1.9, which is close to L1206, HH288, and I\,20343, while I\,21307, I\,22198, and I\,23385 show higher values. 
On the other hand, the CH$_3$CN/HC$_3$N ratios in all of the IM protostars are higher than that in L1527 ($\sim0.02$), which are expected results because L1527 is deficient in COMs.
Hence, the N-bearing COM is more abundant around the IM protostars than the low-mass WCCC source, whereas the CH$_3$OH abundances show a source dependence.

We summarize the chemical characteristics of IM protostars below:
\begin{enumerate}
    \item The compositions of small carbon-chain species are similar to those in L1527.
    \item The larger carbon-chain species tend to be deficient compared to L1527.
    \item Three IM protostars (L1206, HH288, and I\,20343) show similar CH$_3$OH/HC$_3$N abundance ratio as L1527, whereas two IM sources (I\,21307 and I\,22198) show much higher ratio, which implies that the CH$_3$OH abundances depend on the source characteristics.
    \item CH$_3$CN is more abundant around the IM protostar than L1527.
\end{enumerate}

Next, we compared the HC$_5$N/HC$_3$N abundance ratios to those in high-mass protostellar objects (HMPOs) derived by Q-band observations with the Nobeyama 45 m radio telescope \citep{2018ApJ...854..133T}.
We calculated the HC$_5$N/HC$_3$N towards 14 HMPOs where both species have been detected \citep{2018ApJ...854..133T}.
The average HC$_5$N/HC$_3$N ratio is calculated at 0.3, but there is a large scatter from 0.1 to 1.0.
The HC$_5$N/HC$_3$N ratios in the IM protostars are similar to the minimum and average values of HMPOs.

We obtained the HC$_5$N/HC$_3$N and CH$_3$OH/HC$_3$N abundance ratios towards three MYSOs (G\,12.89+0.49, G\,16.86-2.16, and G\,28.28-0.36) from \citet{Taniguchi2018HC5N/CH3OH}.
These three MYSOs are more physically evolved than HMPOs. 
At the MYSO stage, the HCCC mechanism produces cyanopolyynes efficiently \citep{Taniguchi2023ApJS}.
The MYSO G\,12.89+0.49 is found to be a COMs-rich hot core, whereas G\,28.28-0.36 is a carbon-chain-rich/COMs-poor source.
The HC$_5$N/HC$_3$N ratios were derived to 0.2 towards G\,12.89+0.49 and G\,16.86-2.16, and 0.3 towards G\,28.28-0.36, respectively.
The HC$_5$N/HC$_3$N abundance ratios around the IM protostars are consistent with or slightly lower than those of the MYSOs.

The CH$_3$OH/HC$_3$N ratios in the three MYSOs are 21, 12, and 3 in G\,12.89+0.49, G\,16.86-2.16, and G\,28.28-0.36, respectively.
The ratio in I\,00420 is consistent with those in G\,12.89+0.49 and G\,16.86-2.16 within the error, whereas L1206, HH288, and I\,20343 match with that in G\,28.28-0.36.
The ratios in I\,22198 and I\,21307 are higher than that in G\,12.89+0.49 by a factor of approximately 3 and 7, respectively.

In summary, the HC$_5$N/HC$_3$N ratios in the IM protostars are close to those in HMPOs and MYSOs in single-dish scales.
Since we compared the results obtained by the single-dish telescopes, their emission is dominated by warm envelopes rather than the hot core regions. 
The WCCC mechanism works ubiquitously around protostars with various stellar masses and produces similar chemical compositions of carbon-chain species.

\subsection{Relationship with bolometric luminosity and the HC$_5$N/HC$_3$N abundance ratio}

\begin{figure}[ht]
   \centering
    \includegraphics[bb = 0 10 300 250, scale = 0.9]{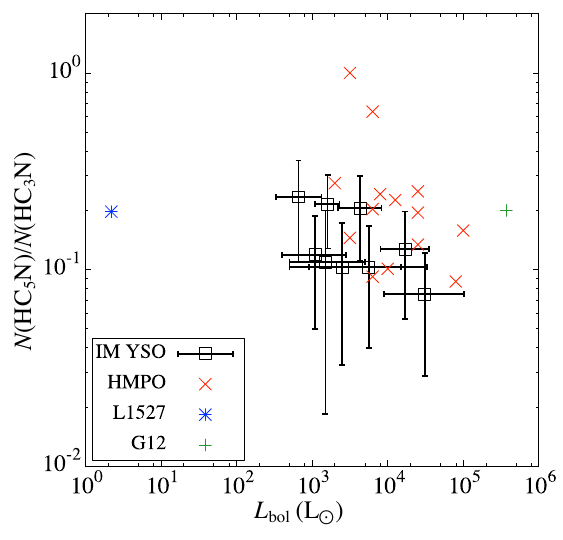}
   \caption{A plot of the HC$_5$N/HC$_3$N column density ratio vs. bolometric luminosity towards various protostars. Their column densities were derived by single-dish observations. Information on the bolometric luminosities of L1527, HMPOs, and G12 are taken from \citet{2002ApJ...575..337S}, \citet{2002ApJ...566..931S}, and \citet{Taniguchi2023ApJS}, respectively.} \label{fig:LvsHC5N}
\end{figure}

Energetic particles such as UV radiation and cosmic rays could enhance the HC$_5$N/HC$_3$N ratios \citep{2017A&A...605A..57F, Taniguchi2019model}.
For instance, \citet{2017A&A...605A..57F} found that the emission of HC$_3$N and HC$_5$N do not coincide in OMC2-FIR4; HC$_3$N emission overlaps relatively well with the continuum emission, whereas HC$_5$N emits only in the eastern-half of it.
In this subsection, we investigate a possible correlation between the bolometric luminosity and the HC$_5$N/HC$_3$N abundance ratio gathering data towards low-mass, IM, and high-mass protostars.

Fig. \ref{fig:LvsHC5N} shows a relationship between the HC$_5$N/HC$_3$N abundance ratios and the source bolometric luminosity.
We plot data towards IM protostars, the low-mass WCCC source L1527 \citep{2019PASJ...71S..18Y}, HMPOs \citep{2018ApJ...854..133T}, and the MYSO G\,12.89+0.49 \citep{Taniguchi2018HC5N/CH3OH} to cover a wide range of bolometric luminosity.
As seen in Fig. \ref{fig:LvsHC5N}, no correlation is found between the bolometric luminosity and the HC$_5$N/HC$_3$N ratio.
The result implies that the formation of cyanopolyynes around protostars is not dominated by the UV radiation and energetic particles.
This ensures that the WCCC mechanism, which depends only on the temperature, forms carbon-chain species.

\subsection{Comparisons of line width and centroid velocity }

\begin{figure*}
   \centering
    \includegraphics[bb = 0 10 600 750, scale = 0.9]{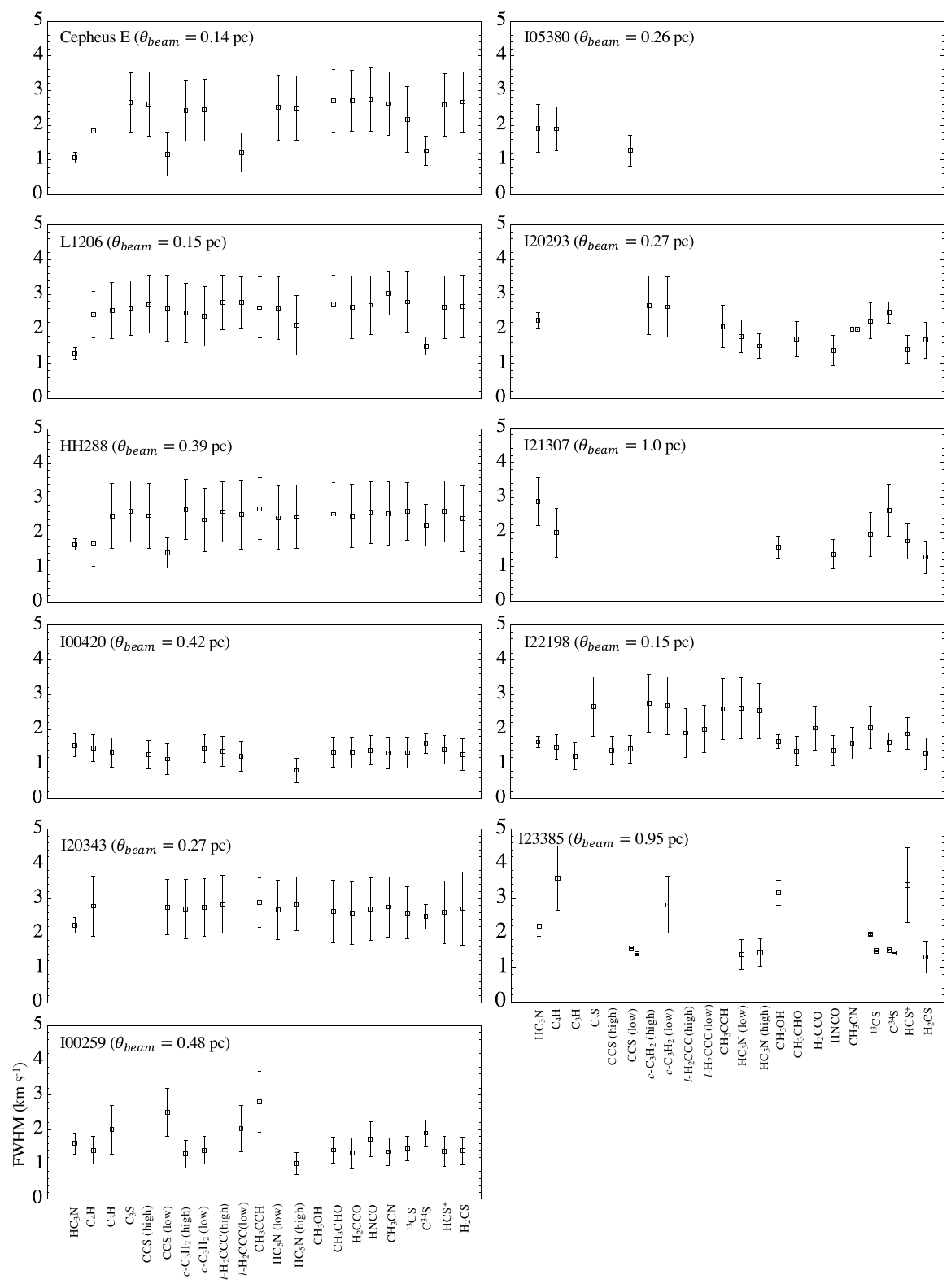}
   \caption{Comparison of line width (FWHM) obtained by the MCMC analyses. $\theta_{beam}$ indicates the linear-scale beam sizes of $40\arcsec$ at each source distance.} \label{fig:FWHM}
\end{figure*}

\begin{figure*}
   \centering
    \includegraphics[bb = 0 10 600 670, scale = 0.98]{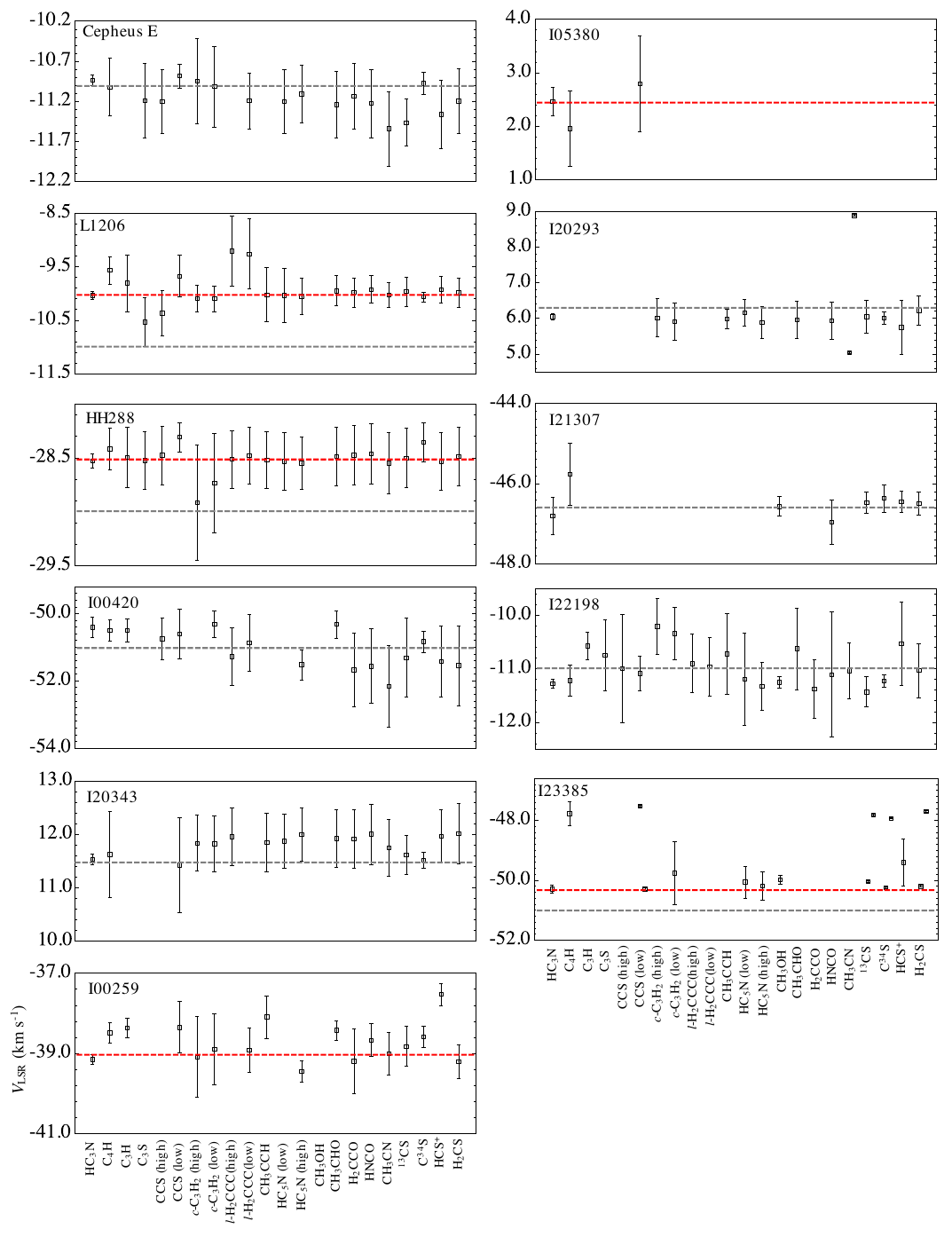}
   \caption{Comparison of the centroid velocity ($V_{\rm {LSR}}$) obtained by the MCMC analyses. The error bars do not include the velocity resolution of spectra ($\approx 0.3$ km\,s$^{-1}$). The gray dashed horizontal lines indicate the systemic velocity of the source (Table \ref{table:sourcelist}). The red dashed horizontal lines indicate the systemic velocity updated or reported based on our results.} \label{fig:Vlsr}
\end{figure*}

\begin{figure}
   \centering
    \includegraphics[bb = 0 15 360 240, scale = 0.68]{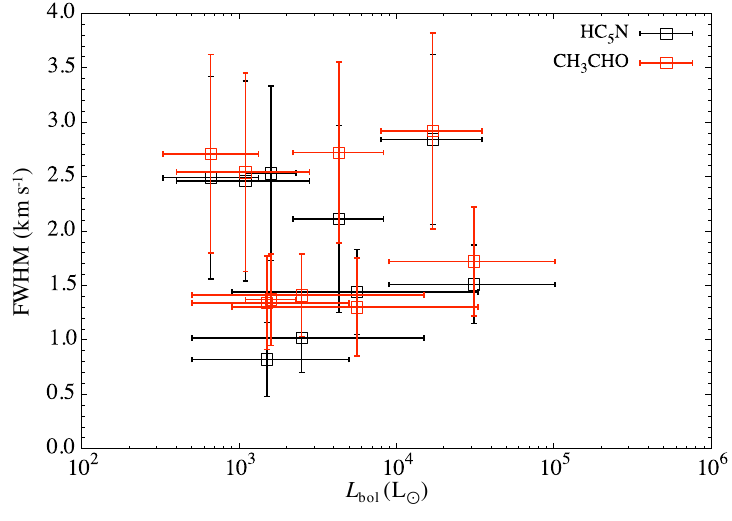}
   \caption{A plot of line widths of HC$_5$N (black) and CH$_3$CHO (red) vs. bolometric luminosity.} \label{fig:LvsFWHM}
\end{figure}

The line width and centroid velocity provide a clue to investigate where each molecular line traces.
In this subsection, we compare these values in each source. 

Fig. \ref{fig:FWHM} shows comparison of line width (FWHM) obtained by the MCMC analyses (Sect. \ref{sec:MCMC}) among different molecular lines for each IM protostar.
We cannot see differences between carbon-chain species, COMs, H$_2$CCO, HNCO, and S-bearing species. 
These results suggest that both carbon-chain species and COMs possibly trace similar regions around the IM protostars, $i.e.$, warm envelopes. 

In Cepheus E, the lines of HC$_3$N, the low upper-state-energy lines of CCS and $l$-H$_2$CCC, and C$^{34}$S show narrower line widths compared to the other lines.
We can see similar trends in HC$_3$N and C$^{34}$S towards L1206, and the low upper-state-energy line of CCS in HH288.
These lines may trace mainly cold envelopes. 
However, these trends are not universal for all of the IM protostars.
The different linear-scale beam sizes ($\approx0.14-1.0$ pc), in other words, different source distances ($\approx0.7-5$ kpc; see Table \ref{table:sourcelist}), may affect these results. 

Fig. \ref{fig:Vlsr} indicates comparison of the centroid velocity  ($V_{\rm {LSR}}$) of each molecular line obtained by the MCMC analyses (Sect. \ref{sec:MCMC}).
The gray dashed lines in each panel, except for I\,00259 and I\,05380, indicate the systemic source velocities previously reported (Table \ref{table:sourcelist}).
All of the lines show almost similar values as the source systemic velocities in Cepheus E, I\,00420, I\,20343, I\,20293, I\,21307, and I\,22198.
Molecular lines, except for $c$-C$_3$H$_2$, in HH288 have slightly higher values than the source systemic velocity.
All of the molecular lines seem to show larger values than the systemic source velocity in L1206.
This could happen because the source velocity in L1206 was derived by the maser.
The thermal molecular lines likely have different velocity components from the maser lines.

We can see velocity shifts in the molecular lines from the source systemic velocities in L1206, HH288, and I\,23385.
There are no available data for the systemic velocities of I\,00259 and I\,05380. 
Here, we revise and provide systemic velocities of these protostars based on the results of HC$_3$N, which is a good dense core tracer; -10 km\,s$^{-1}$ for L1206, -28.5 km\,s$^{-1}$ for HH288, -39 km\,s$^{-1}$ for I\,00259, 2.5 km\,s$^{-1}$ for I\,05380, and -50.3 km\,s$^{-1}$ for I\,23385, respectively.
Table \ref{table:vlsr} in Appendix \ref{sec:append1} summarizes this information.

Regarding CH$_3$CN in I\,20293 and $^{13}$CS and C$^{34}$S in I\,23385, two velocity components were identified.
The two velocity components of CH$_3$CN in I\,20293 are different from the other lines, but the lower velocity component is marginally consistent with HCS$^+$ within their errors.
We cannot identify the cause(s) of the velocity shifts in the single-dish observations.
In the case of the isotopomers of CS in I\,23385, the low-velocity components ($\sim-50$ km\,s$^{-1}$) are similar to most of the other molecular lines, whereas the high-velocity component is similar to those of C$_4$H and the high-velocity component of CCS (low). 
As seen in Fig. \ref{fig:FWHM}, the C$_4$H lines in I\,23385 show wider line features ($\sim 3.6$ km\,s$^{-1}$) compared to the other sources.
Hence, the emission region of C$_4$H in I\,23385 may be different from the other IM protostars, $e.g.$, cavity wall of molecular outflows.
Such a difference may suggest that I\,23385 contains a more massive star than IM protostars and the powerful outflow(s) affect the spatial distributions of these molecules, which agrees with \citet{2023A&A...673A.121B}.

Fig. \ref{fig:LvsFWHM} shows a plot of line widths of HC$_5$N (black) and CH$_3$CHO (red) vs. the bolometric luminosity.
There is no correlation between the bolometric luminosity and line widths of both species in the bolometric luminosity range of the target IM protostars.
These results may imply that the observed lines trace regions less affected by the central stars; no correlation is caused by the low-angular resolution data obtained by the single-dish telescope.
Similar no correlations between the bolometric luminosity and line widths of carbon-chain species were found towards HMPOs obtained with the Nobeyama 45 m telescope \citep{Taniguchi2019HMPOsurvey}.

\subsection{The $cyclic$-to-$linear$ ratio of the C$_3$H$_2$ isomer}

\begin{figure*}
   \centering
    \includegraphics[bb = 0 10 600 220, scale = 0.65]{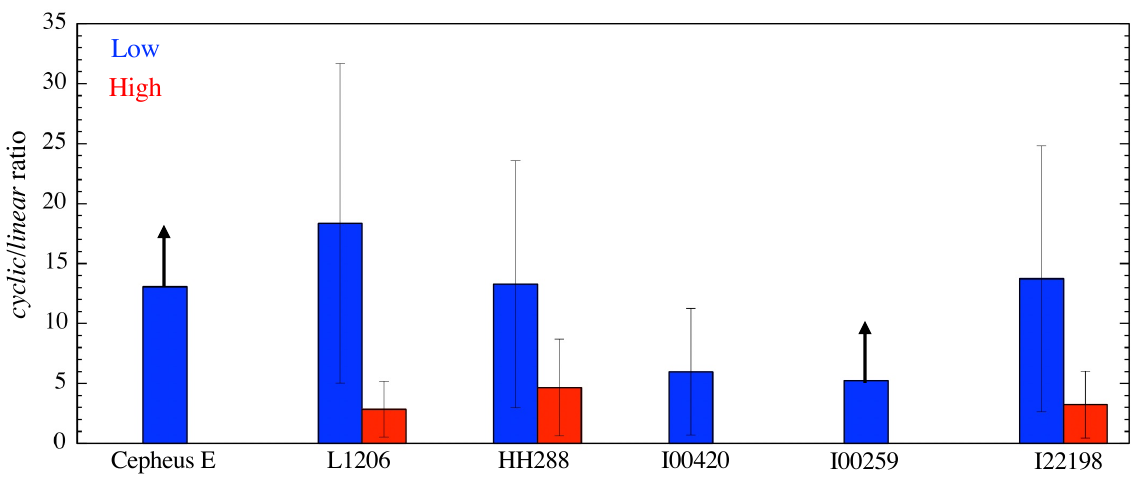}
   \caption{The $cyclic$-to-$linear$ ratios of C$_3$H$_2$ isomers, $c$-C$_3$H$_2$ and $l$-H$_2$CCC. The blue and red bars indicate the ratios for the low and high-temperature components depending on the upper state energies (see Sect. \ref{sec:MCMC}).} \label{fig:clratio}
\end{figure*}

The $cyclic$-to-$linear$ ratios (hereafter $c/l$ ratio) have been investigated by combining observations and theoretical studies \citep{2016A&A...591L...1S, 2017MNRAS.470.4075L}.
Physical conditions likely affect the $c/l$ ratios.
For instance, the $c/l$ ratios of C$_3$H$_2$ were found to be $110\pm30$ and $30\pm10$ for molecular clouds with densities of $10^4$ cm$^{-3}$ and $4\times10^5$ cm$^{-3}$, respectively \citep{2017MNRAS.470.4075L}.
The $c/l$ ratio at the starless clump in the Serpens South cluster-forming region was derived at $58\pm6$ \citep{2024arXiv240916492T}.
These results imply that the density is a key factor to produce differences in the $c/l$ ratio of the C$_3$H$_2$ isomers.
It was proposed that isomerization reactions of ``$l$-C$_3$H$_2$ + H $\rightarrow$ $c$-C$_3$H$_2$ + H'' and ``$t$-C$_3$H$_2$ + H $\rightarrow$ $c$-C$_3$H$_2$ + H'' are important for the high $c/l$ ratio in low-density conditions \citep{2017MNRAS.470.4075L}. 
In this subsection, we compare the $c/l$ ratio of the C$_3$H$_2$ isomers among the IM protostars.

Fig. \ref{fig:clratio} shows a comparison of the $c/l$ ratios of the C$_3$H$_2$ isomers, $c$-C$_3$H$_2$ and $l$-H$_2$CCC. 
The ``Low'' (blue) and ``High'' (red) mean that the ratios were derived using the low $E_{\rm{up}}$ lines assuming the excitation temperature of 10 K and the high $E_{\rm{up}}$ lines assuming the excitation temperature of 20 K, respectively (see Sect. \ref{sec:ana}).
Since $l$-H$_2$CCC has been detected tentatively in Cepheus E and I\,00259 and their column densities are the upper limits, their $c/l$ ratios are the lower limits. Spectra of $c$-C$_3$H$_2$ show weak peak intensities in I\,00420 and its relative error is large. If we exclude these three sources with large uncertainties, the Low components show the $c/l$ ratio of a range of 10--20. 
The Low components always show higher ratios compared to the High components ($\sim3-5$) in the three sources in which both of the components have been detected, even though there are large errors.

This may reflect that outer cold envelopes (the Low component) have lower densities than inner warm regions where the WCCC mechanism occurs (the High component).
However, there is a possibility that the temperature may affect the $c/l$ ratio. 
Since the previous theoretical studies did not consider the warm-up phase, this point is still unclear.

The $c$/$l$ ratio in the WCCC source is lower than those in the cold prestellar cores.; the $c/l$ ratios in prestellar cores were derived to be $\sim30-110$ \citep{2016A&A...591L...1S, 2017MNRAS.470.4075L}, whereas the ratio in L1527 was derived to be 12 \citep{2016A&A...591L...1S}.
Such a tendency can be regarded in Fig. \ref{fig:clratio}; the Low components show higher values than those of the High components.
These two different $c/l$ ratios support the scenario that carbon-chain species exist in both outer less-dense envelopes and inner denser envelopes.

\subsection{Carbon and sulfur isotopic ratios in CS}

\begin{figure*}
   \centering
    \includegraphics[bb = 0 10 550 270, scale = 0.72]{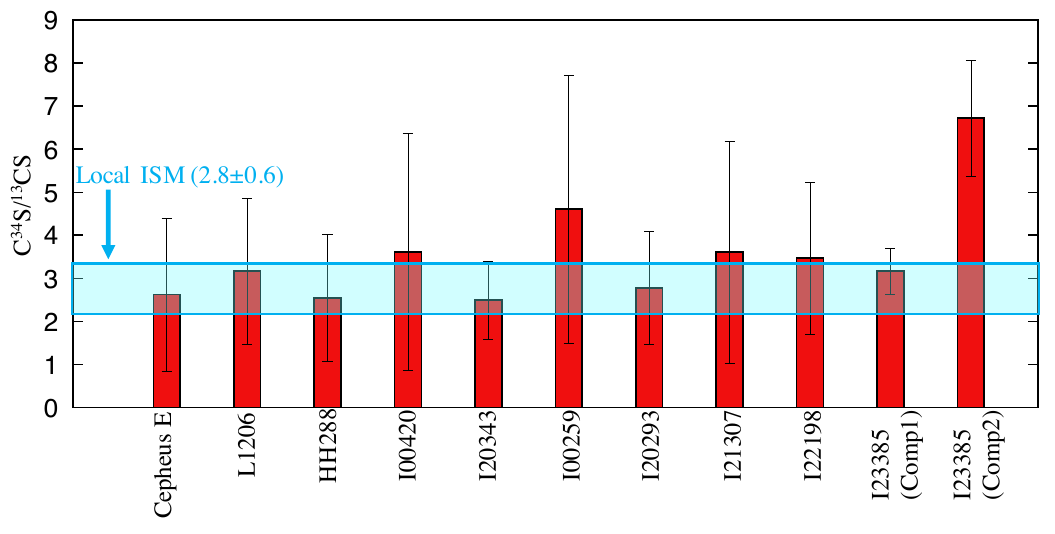}
   \caption{C$^{34}$S/$^{13}$CS ratios among the 10 IM protostars. Comp\,1 and Comp\,2 of I\,23385 correspond to the velocity components of -50 km\,s$^{-1}$ and -47.8 km\,s$^{-1}$, respectively. The local ISM value ($2.8\pm0.6$) was calculated from the results of \citet{2023A&A...670A..98Y}.} \label{fig:Sisotope}
\end{figure*}

Fig. \ref{fig:Sisotope} shows a comparison of the C$^{34}$S/$^{13}$CS abundance ratios among the 10 protostars. 
The light blue region indicates the local ISM value \citep[$2.8\pm0.6$;][]{2023A&A...670A..98Y}, which was calculated using the following formulae adopting their results of CS isotopologues; 
\begin{equation} \label{equ:isoratio}
\frac{\mathrm{C}^{34}\mathrm{S}}{^{13}\mathrm{CS}} = \frac{^{12}\mathrm{C}}{^{13}\mathrm{C}} \times \frac{^{34}\mathrm{S}}{^{32}\mathrm{S}}.
\end{equation}

The observed ratios are consistent with the local ISM value within the errors.
The Comp\,2 of I\,23385, whose velocity component is $V_{\rm {LSR}}\approx-47.8$ km\,s$^{-1}$, has a higher value than that of the local ISM.
Since I\,23385 shows the highest bolometric luminosity and is a high-mass protostar, one possible scenario to produce such isotope anomaly is that the local UV radiation destroys less isotopologue ($^{13}$CS) more efficiently, $i.e.,$ the self-shielding effect.
Or, the PDR-like chemistry at the cavity wall of the molecular outflow may affect the chemistry of the Comp\,2.
Three outflows or jets have been identified in this high-mass protostellar system by several shock tracers such as SiO, H$_2$, [Fe II], and [Ne II] \citep{2023A&A...673A.121B}.
If Comp\,2 traces the outflow components, the self-shielding effect may work and produce the high C$^{34}$S/$^{13}$CS ratio only in Comp\,2.
High-angular-resolution observations to resolve the two different components are necessary to reveal the origin of the isotope anomaly.

\section{Conclusions} \label{sec:con}

We have conducted Q-band line survey observations towards 11 protostars, which are selected from the sub-sample source list of the SOMA project, with the Yebes 40 m telescope.
The main findings and conclusions of this paper are as follows.

\begin{enumerate}

    \item We have detected nine carbon-chain species (HC$_3$N, HC$_5$N, C$_3$H, C$_4$H, $linear-$H$_2$CCC, $cyclic-$C$_3$H$_2$, CCS, C$_3$S, and CH$_3$CCH), three COMs (CH$_3$OH, CH$_3$CHO, CH$_3$CN), H$_2$CCO, HNCO, and four S-bearing species ($^{13}$CS, C$^{34}$S, HCS$^+$m and H$_2$CS) from the 11 protostars. 
    
    \item The derived rotational temperatures of HC$_5$N are approximately 20 -- 30 K, suggesting that carbon-chain molecules exist in warm regions around the IM protostars. The rotational temperatures are consistent with those derived in low-mass and high-mass protostars. We need to observe high $J$ lines to clarify the presence of hot components.

    \item Based on the comparisons of the chemical compositions around the IM protostars to those in the low-mass WCCC source L1527 and HMPOs/MYSO, the HC$_5$N/HC$_3$N ratios are found to be similar to those around low-mass and high-mass protostars. Since the beam size of the single-dish telescope is much larger than the hot regions around IM protostars, the detected carbon-chain emission comes from warm envelopes where the WCCC mechanism is dominant. To confirm the HCCC mechanism, we need interferometric observations to derive their spatial distributions.

    \item No correlations have been found between the bolometric luminosity and the HC$_5$N/HC$_3$N abundance ratio and line width. This implies that these cyanopolyynes are not formed by the PDR chemistry, and supports the WCCC scenario.

    \item The $c/l$ ratios of the C$_3$H$_2$ isomers suggest that these species exist in regions with at least two different physical conditions; the less dense outer regions and denser inner regions. These results support our assumption that carbon-chain species exist in outer cold envelopes too.

    \item The C$^{34}$S/$^{13}$CS ratios in the IM protostars generally agree with the value in the local ISM. However, the second velocity component in I\,23385 ($\sim -47$ km\,s$^{-1}$) shows a higher value. Since this is a high-mass protostar with a high bolometric luminosity, the enhancement of the local UV radiation may produce such an isotopic anomaly. Or, the PDR-like chemistry at the cavity wall may affect it.

\end{enumerate}

Our results confirm that carbon-chain species form in warm gas around the IM protostars, and the presence of the WCCC mechanism is robust.
Future interferometric observations and higher frequency line survey observations will further constrain the chemical compositions and carbon-chain formation mechanisms around IM protostars, $i.e.,$ confirmation of the HCCC mechanism and comparisons of spatial distributions between carbon-chain species and COMs.

\begin{acknowledgements}
We deeply appreciate the staff of the Radiotelescope Administration of the Yebes Observatory (RYAO).
K.T. is supported by JSPS KAKENHI grant Nos. JP20K14523, 21H01142, 24K17096, and 24H00252. JCT acknowledges support from ERC Advanced Grant 788829 (MSTAR).
M.G.-G. is partially supported by the research grant Nebulaeweb/ eVeNts (PID2019-105203GB-C21) of the Spanish AEI(MICIU).
R.F. acknowledges support from the grants Juan de la Cierva FJC2021-046802-I, PID2020-114461GB-I00, PID2023-146295NB-I00, and CEX2021-001131-S funded by MCIN/AEI/ 10.13039/501100011033 and by ``European Union NextGenerationEU/PRTR''.
Y.-L.Y. acknowledges support from Grant-in-Aid from the Ministry of Education, Culture, Sports, Science, and Technology of Japan (20H05845, 20H05844, 22K20389), and a pioneering project in RIKEN (Evolution of Matter in the Universe). 
We thank the anonymous referee whose comments helped improve the paper.
\end{acknowledgements}

%
%

\bibliographystyle{aa}
\bibliography{References}{}

\begin{appendix} 

\section{Spectra and fitting results} \label{sec:append1}

\begin{table}
\caption{Information on molecular lines}            
\label{table:linesum}   
\begin{center}    
\scalebox{0.85}{
\begin{tabular}{l l c c}       
\hline\hline               
Species & Transition & Frequency$^{(a)}$ & $E_{\rm {up}}/k$ \\
        &            &  (GHz)         & (K) \\
\hline                        
HC$_3$N & $4-3$ & 36.392324 & 4.4 \\
HC$_3$N & $5-4$ & 45.490314 & 6.5 \\
HC$_5$N & $12-11$ & 31.951772 & 9.9 \\
HC$_5$N & $13-12$ & 34.614387 & 11.6 \\
HC$_5$N & $14-13$ & 37.276994 & 13.4 \\
HC$_5$N & $15-14$ & 39.939591 & 15.3 \\
HC$_5$N & $16-15$ & 42.602153 & 17.4 \\
HC$_5$N & $17-16$ & 45.264720 & 19.6 \\
HC$_5$N & $18-17$ & 47.927275 & 21.9 \\
C$_3$H  & $J=\frac{3}{2}-\frac{1}{2}, \Omega=\frac{1}{2}, F=2-1, l=f$ & 32.627297 & 1.6 \\
C$_3$H  & $J=\frac{3}{2}-\frac{1}{2}, \Omega=\frac{1}{2}, F=2-1, l=e$ & 32.660645 & 1.6 \\
C$_4$H & $N=4-3, J=\frac{9}{2}-\frac{7}{2}, F=4-3$$^{(b)}$ & 38.049616 & 4.6 \\
C$_4$H & $N=4-3, J=\frac{9}{2}-\frac{7}{2}, F=5-4$$^{(b)}$ & 38.049691 & 4.6 \\
C$_4$H & $N=4-3, J=\frac{7}{2}-\frac{5}{2}, F=4-3$$^{(b)}$ & 38.088441  & 4.6 \\
C$_4$H & $N=4-3, J=\frac{7}{2}-\frac{5}{2}, F=3-2$$^{(b)}$ & 38.088481 & 4.6 \\
C$_4$H & $N=5-4, J=\frac{11}{2}-\frac{9}{2}, F=5-4$$^{(b)}$ & 47.566770 & 6.8 \\
C$_4$H & $N=5-4, J=\frac{11}{2}-\frac{9}{2}, F=6-5$$^{(b)}$ & 47.566814 & 6.8 \\
C$_4$H & $N=5-4, J=\frac{9}{2}-\frac{7}{2}, F=5-4$$^{(b)}$ & 47.605490 & 6.9 \\
C$_4$H & $N=5-4, J=\frac{9}{2}-\frac{7}{2}, F=4-3$$^{(b)}$ & 47.605502 & 6.9 \\
$l$-H$_2$CCC & $2_{1,2}-1_{1,1}$ & 41.198335 & 16.3  \\
$l$-H$_2$CCC & $2_{0,2}-1_{0,1}$ & 41.584676 & 3.0 \\
$l$-H$_2$CCC & $2_{1,1}-1_{1,0}$ & 41.967671 & 16.4 \\
$c$-C$_3$H$_2$ & $3_{2, 1}-3_{1,2}$ & 44.104777 & 18.2 \\
$c$-C$_3$H$_2$ & $2_{1,1}- 2_{0,2}$ & 46.755610 & 8.7 \\
CCS & $J_N= 3_2-2_1$ & 33.751370 & 3.2 \\
CCS & $J_N= 3_3-2_2$ & 38.866420 & 12.4 \\
CCS & $J_N= 3_4-2_3$ & 43.981019 & 12.9 \\
CCS & $J_N= 4_3-3_2$ & 45.379046 & 5.4 \\
C$_3$S & $6-5$ & 34.684369 & 5.8 \\
C$_3$S & $7-6$ & 40.465015 & 7.8 \\
C$_3$S & $8-7$ & 46.245624 & 10.0 \\
CH$_3$CCH & $2_1-1_1$ & 34.182760 & 9.7 \\
CH$_3$CCH & $2_0-1_0$ & 34.183414 & 2.5 \\
CH$_3$OH & $4_{1,4}-3_{-0,3}$ $E$ & 36.169261 & 28.8  \\
CH$_3$OH & $7_{0,7}-6_{1,6}$ $A$ & 44.069367 & 65.0 \\
CH$_3$OH & $1_{0,1}-0_{0,0}$ $A$ & 48.372460 & 2.3 \\
CH$_3$OH & $1_{-0,1}-0_{-0,0}$ $E$ & 48.376887 & 15.4 \\
CH$_3$CHO & $2_{0,2}-1_{0,1}$ $E$ & 38.506035  & 2.9 \\
CH$_3$CHO & $2_{0,2}-1_{0,1}$ $A$ & 38.512079 & 2.8 \\
CH$_3$CHO & $2_{1,1}-1_{1,0}$ $E$ & 39.362537 & 5.2 \\
H$_2$CCO & $2_{1,2}-1_{1,1}$ & 40.039022 & 15.9 \\
H$_2$CCO & $2_{1,1}-1_{1,0}$ & 40.793832 & 16.0 \\
HNCO & $2_{0,2}-1_{0,1}$ & 43.962996 & 3.2  \\
CH$_3$CN & $2_1-1_1$ & 36.794765 & 9.8 \\
CH$_3$CN & $2_0-1_0$ & 36.795475 & 2.6 \\
$^{13}$CS & $1-0$ & 46.247563 & 2.2 \\
C$^{34}$S & $1-0$ & 48.206941 & 2.3 \\
HCS$^+$ & $1-0$ & 42.674195 & 2.0 \\
H$_2$CS & $1_{0,1}-0_{0,0}$ & 34.351430 & 1.6 \\
\hline                                  
\end{tabular}
}
\end{center}
\tablefoot{(a) Taken from the Cologne Database for Molecular Spectroscopy \citep[CDMS,][]{CDMS2016} except for CH$_3$CHO whose values are taken from the JPL catalog \citep{JPL1998}. (b) These lines are blended with the closest lines.}
\end{table}

\begin{figure*}[ht]
   \centering
    \includegraphics[bb = 0 15 550 720, scale = 0.9]{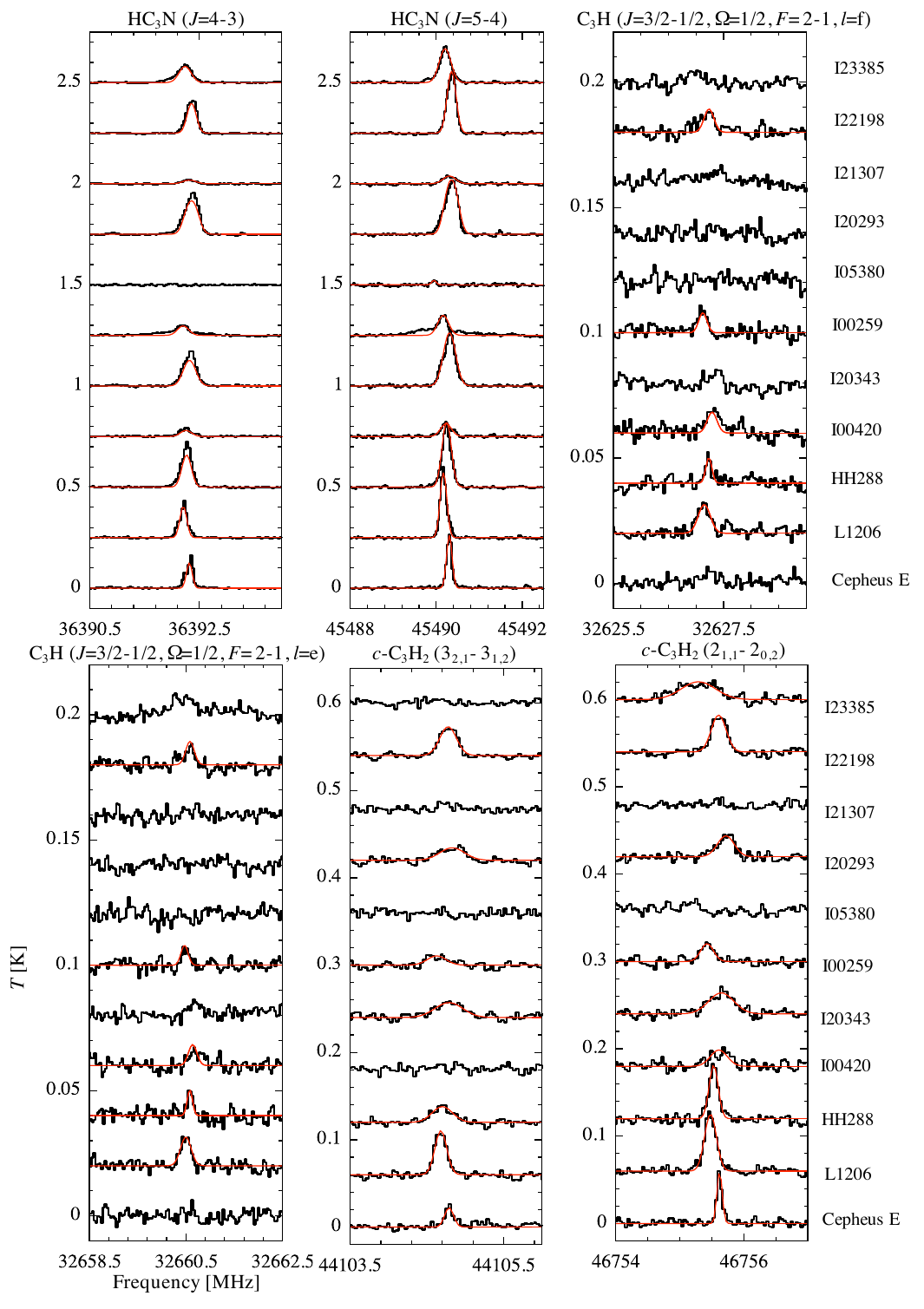}
   \caption{Spectra of three carbon-chain species (HC$_3$N, C$_3$H, $c$-C$_3$H$_2$) towards 11 protostars. Black lines indicate the observational spectra and red curves indicate the best-fitting models with the MCMC method. The order of sources is the same in all of the panels.} \label{fig:speccarbon3}
\end{figure*}

\begin{figure*}[ht]
   \centering
    \includegraphics[bb = 0 15 550 670, scale = 0.9]{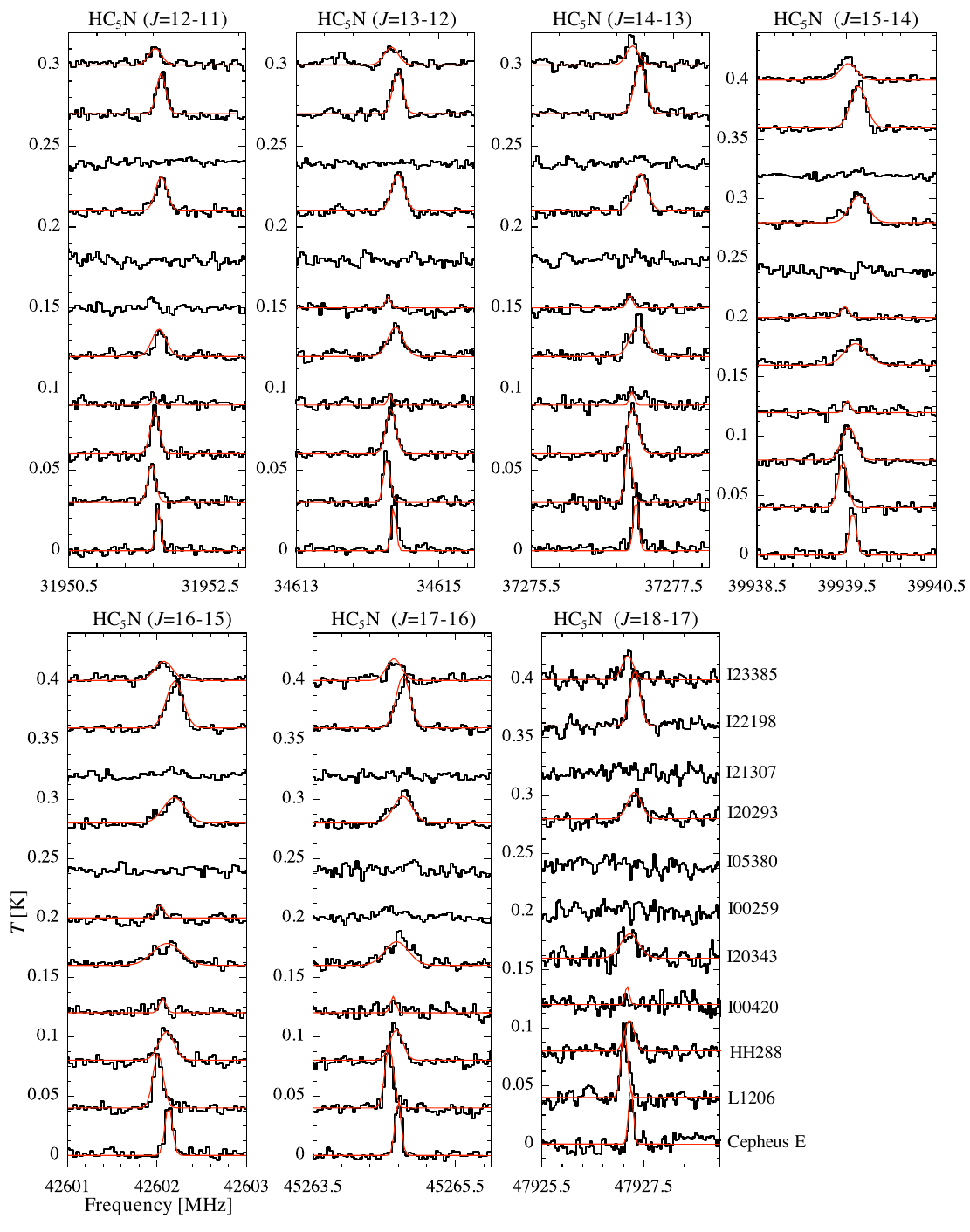}
   \caption{Spectra of HC$_5$N towards 11 protostars. Black lines indicate the observational spectra and red curves indicate the best-fitting models with the MCMC method. The order of sources is the same in all of the panels.} \label{fig:specHC5N}
\end{figure*}

\begin{figure*}[ht]
   \centering
    \includegraphics[bb = 0 15 550 690, scale = 0.9]{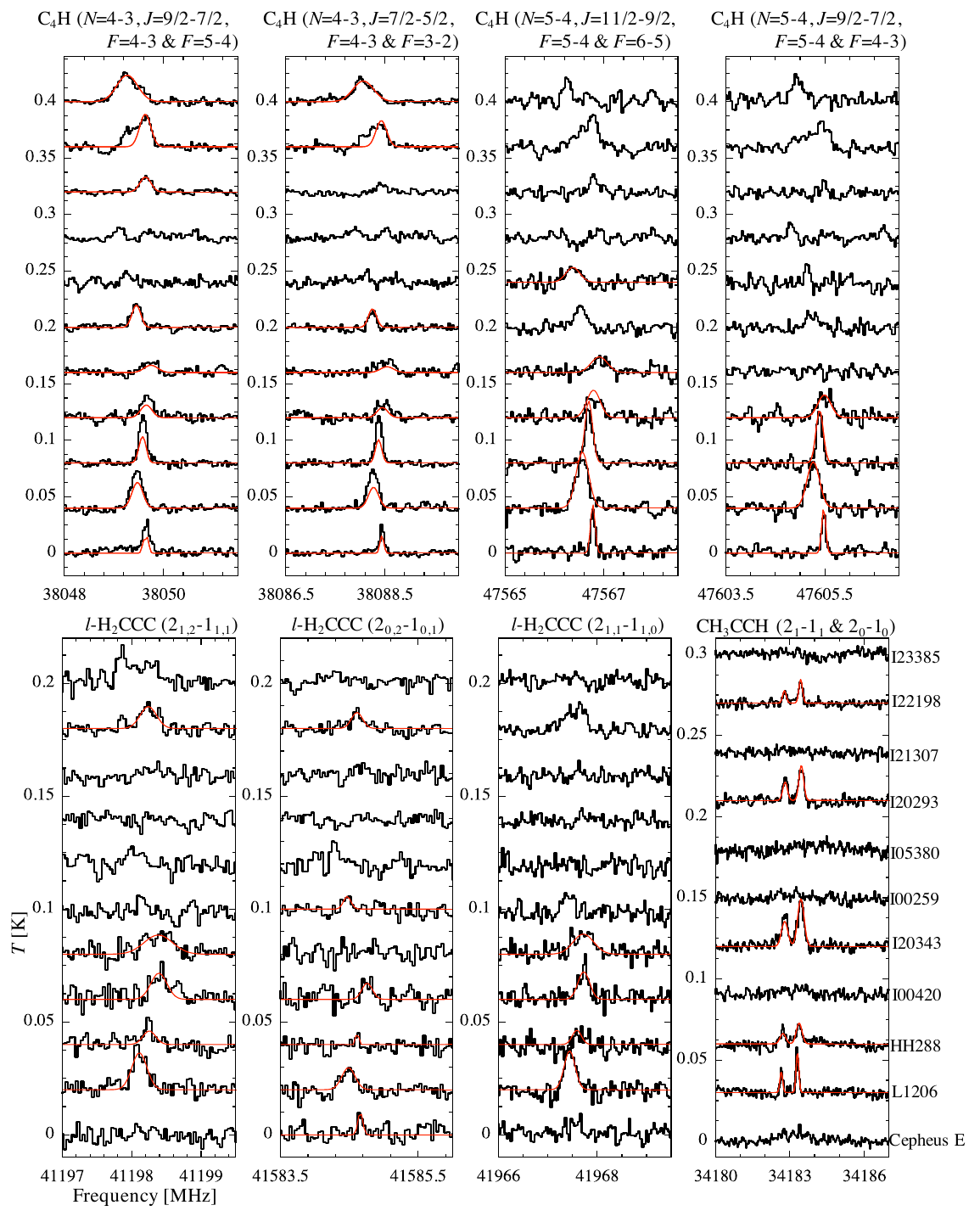}
   \caption{Spectra of C$_4$H, $l$-H$_2$CCC, and CH$_3$CCH towards 11 protostars. Black lines indicate the observational spectra and red curves indicate the best-fitting models with the MCMC method. The order of sources is the same in all of the panels.} \label{fig:speccarbon4}
\end{figure*}

\begin{figure*}[ht]
   \centering
    \includegraphics[bb = 0 15 550 670, scale = 0.9]{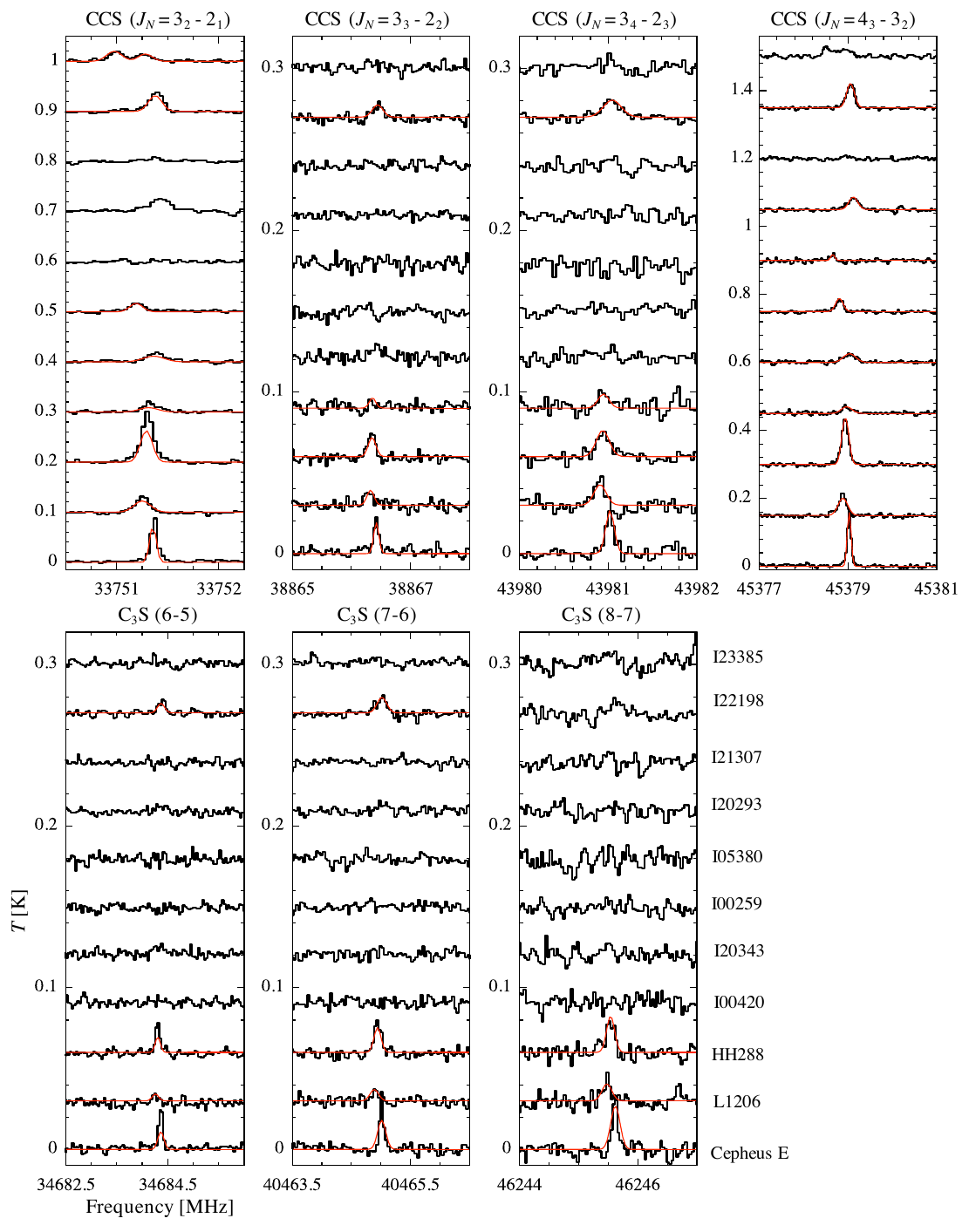}
   \caption{Spectra of CCS and C$_3$S towards 11 protostars. Black lines indicate the observational spectra and red curves indicate the best-fitting models with the MCMC method. The order of sources is the same in all of the panels.} \label{fig:specccs}
\end{figure*}

\begin{figure*}[ht]
   \centering
    \includegraphics[bb = 0 10 520 520, scale = 0.95]{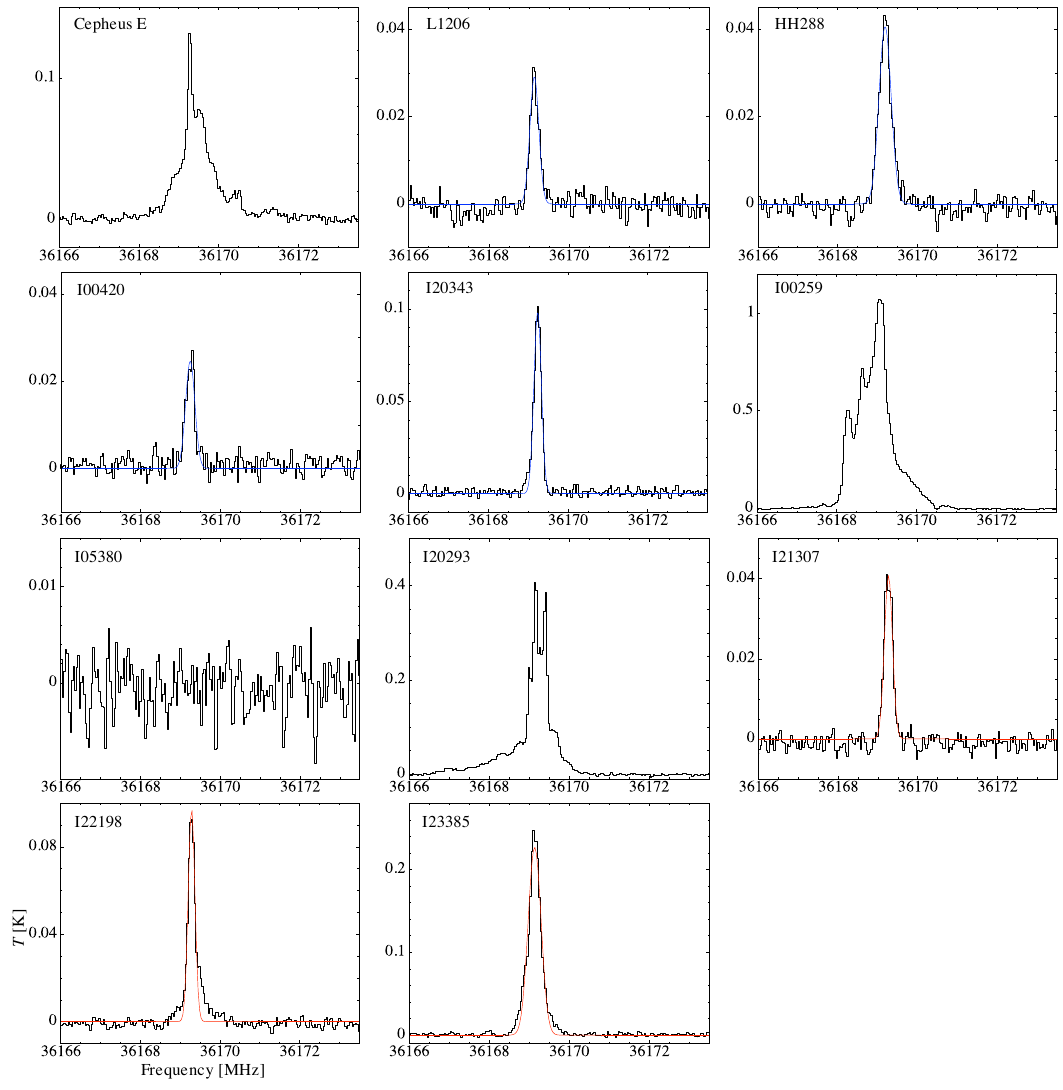}
   \caption{Spectra of CH$_3$OH ($4_{1,4}-3_{-0,3}$ $E$) towards 11 protostars. Black lines indicate the observational spectra and red curves indicate the best-fitting models with the MCMC method. Blue curves show the results of the Gaussian fitting. Note that L1206, HH\,288, I\,00420, I\,20343, and I\,22198 were analyzed by the rotational diagram method.} \label{fig:specCH3OH1}
\end{figure*}

\begin{figure*}[ht]
   \centering
    \includegraphics[bb = 0 10 520 520, scale = 0.95]{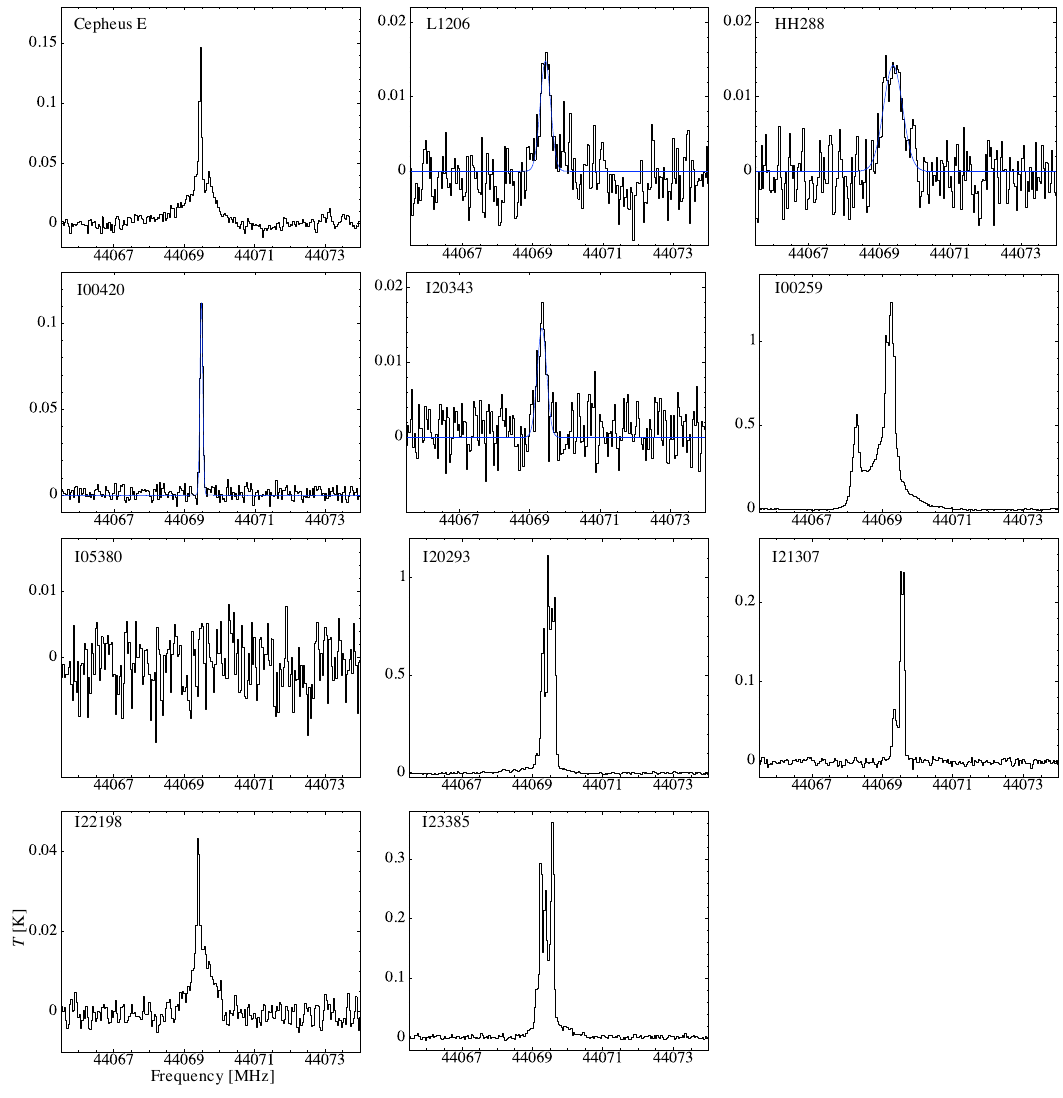}
   \caption{Spectra of CH$_3$OH ($7_{0,7}-6_{1,6}$ $A$) towards 11 protostars. Black lines indicate the observational spectra. Blue curves show the results of the Gaussian fitting. Note that L1206, HH\,288, I\,00420, I\,20343, and I\,22198 were analyzed by the rotational diagram method.} \label{fig:specCH3OH2}
\end{figure*}

\begin{figure*}[ht]
   \centering
    \includegraphics[bb = 0 10 520 520, scale = 0.95]{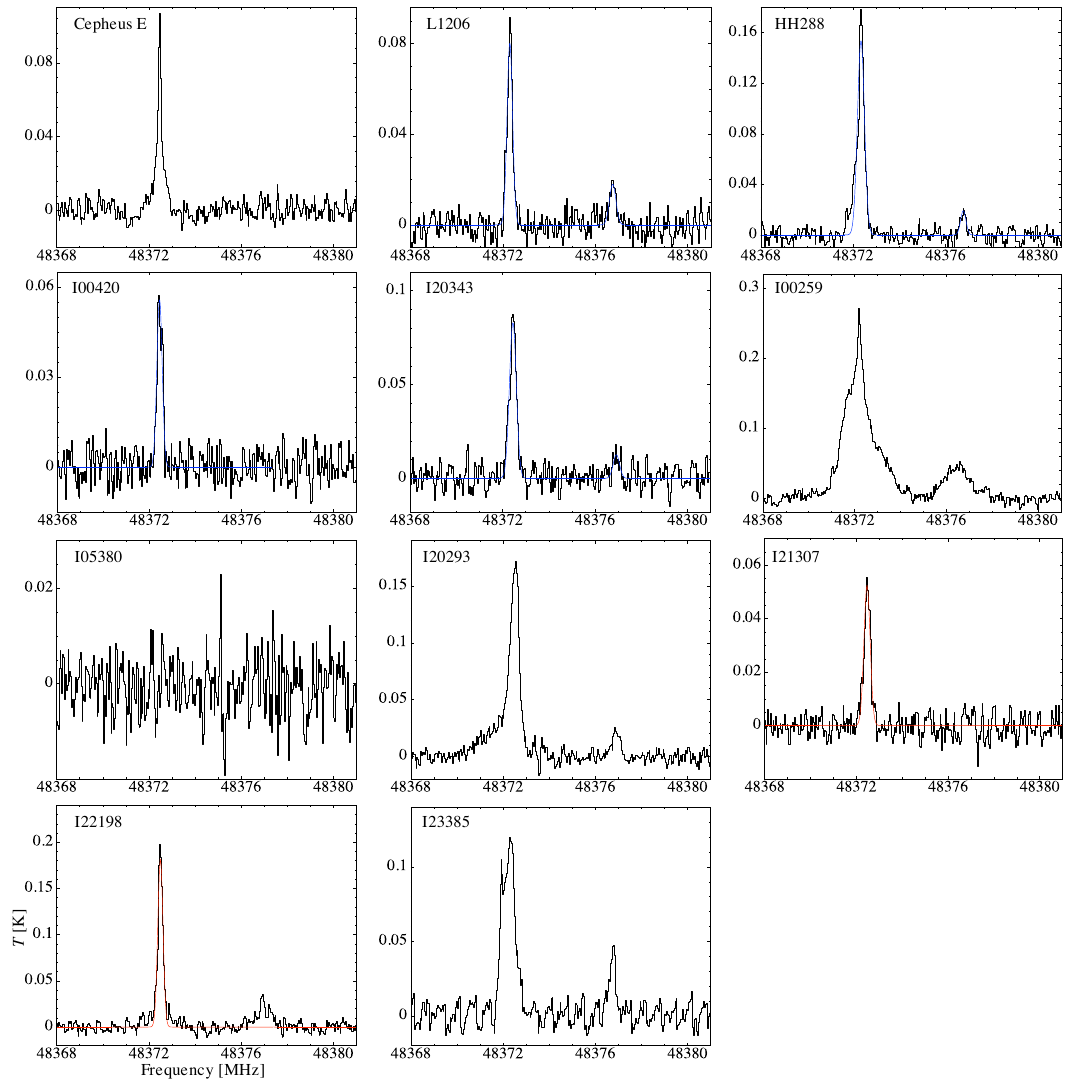}
   \caption{Spectra of CH$_3$OH ($1_{0,1}-0_{0,0}$ $A$ and $1_{-0,1}-0_{-0,0}$ $E$) towards 11 protostars. Black lines indicate the observational spectra and red curves indicate the best-fitting models with the MCMC method. Blue curves show the results of the Gaussian fitting. Note that L1206, HH\,288, I\,00420, I\,20343, and I\,22198 were analyzed by the rotational diagram method. The standing features are seen in the spectra of I\,23385 and we did not analyze this line.} \label{fig:specCH3OH3}
\end{figure*}

\begin{figure*}[ht]
   \centering
    \includegraphics[bb = 0 15 500 720, scale = 0.9]{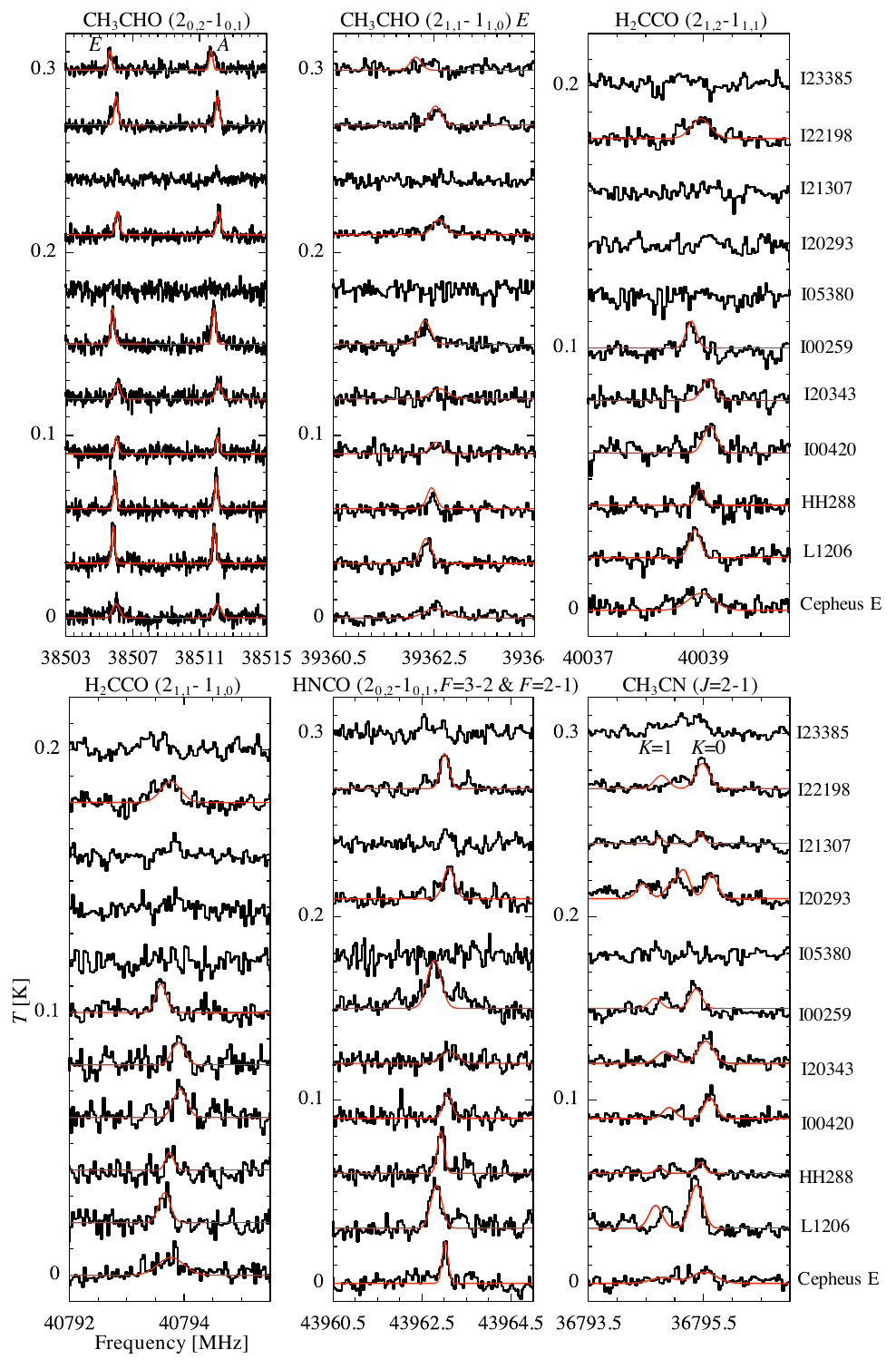}
   \caption{Spectra of CH$_3$CHO, H$_2$CCO, HNCO, and CN$_3$CN, towards 11 protostars. Black lines indicate the observational spectra and red curves indicate the best-fitting models with the MCMC method. The order of sources is the same in all of the panels.} \label{fig:specCOM}
\end{figure*}

\begin{figure*}[ht]
   \centering
    \includegraphics[bb = 0 5 350 760, scale = 0.85]{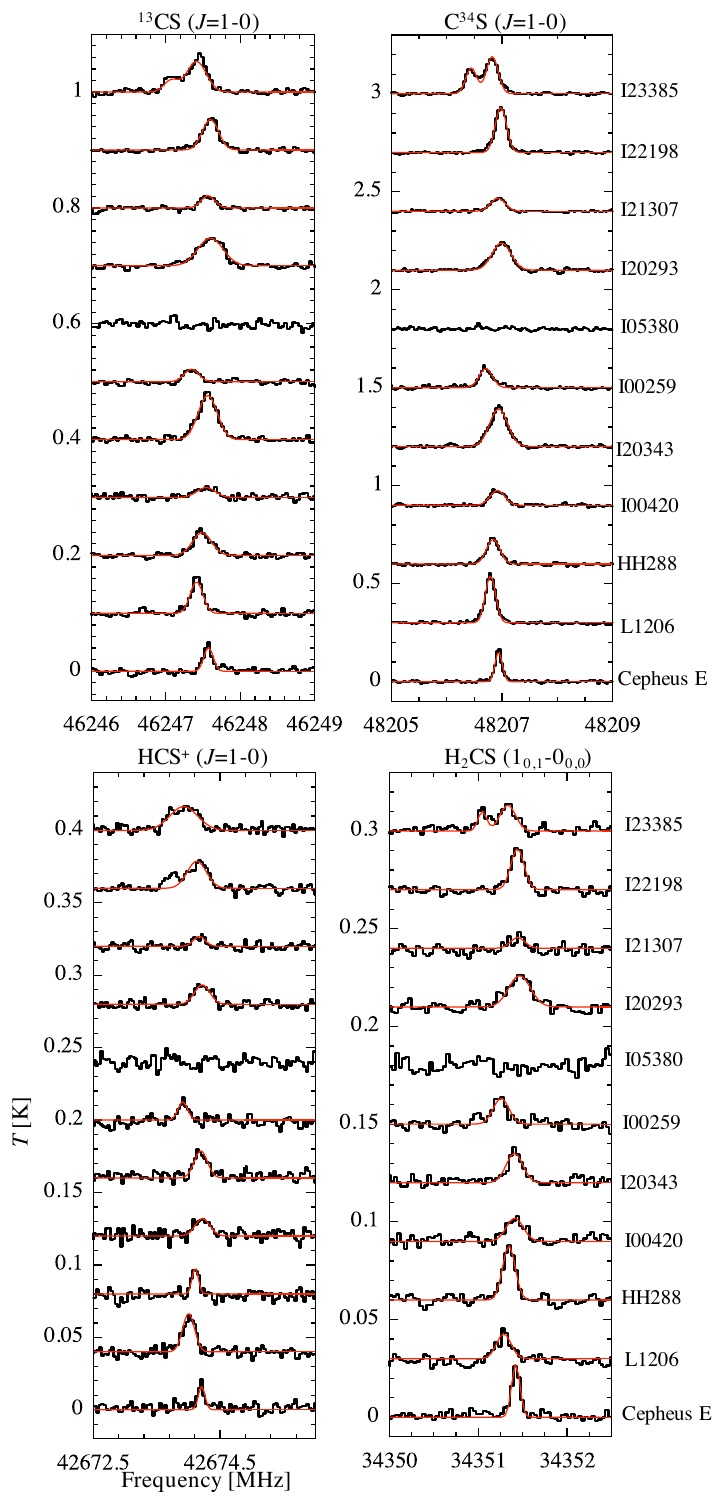}
   \caption{Spectra of simple S-bearing species towards 11 protostars. Black lines indicate the observational spectra and red curves indicate the best-fitting models with the MCMC method. The order of sources is the same in all of the panels.} \label{fig:specS}
\end{figure*}

Figs. \ref{fig:speccarbon3}--\ref{fig:specS} present all of the spectra analyzed in this paper. 
Black lines and red curves indicate the observed spectra and the best-fitting model with the MCMC method.
Information on these lines is summarized in Table \ref{table:linesum}.

Table \ref{table:vlsr} summarizes the line width (FWHM) and the velocity component ($V_{\rm {LSR}}$) obtained by the MCMC analysis.

\begin{sidewaystable*}
\caption{Summary of line widths and centroid velocity.}\label{table:vlsr}
\scalebox{0.5}{
\begin{tabular}{lcccccccccccccccccccccc} 
\hline\hline             
Species & \multicolumn{2}{c}{Cepheus E} & \multicolumn{2}{c}{L1206} & \multicolumn{2}{c}{HH288} & \multicolumn{2}{c}{I\,00420} & \multicolumn{2}{c}{I\,20343} & \multicolumn{2}{c}{I\,00259} & \multicolumn{2}{c}{I\,05380} & \multicolumn{2}{c}{I\,20293} & \multicolumn{2}{c}{I\,21307} & \multicolumn{2}{c}{I\,22198} & \multicolumn{2}{c}{I\,23385} \\
 & FWHM  & $V_{\rm {LSR}}$ & FWHM  & $V_{\rm {LSR}}$ & FWHM  & $V_{\rm {LSR}}$ & FWHM  & $V_{\rm {LSR}}$ & FWHM  & $V_{\rm {LSR}}$ & FWHM  & $V_{\rm {LSR}}$ & FWHM  & $V_{\rm {LSR}}$ & FWHM  & $V_{\rm {LSR}}$ & FWHM  & $V_{\rm {LSR}}$ & FWHM  & $V_{\rm {LSR}}$ & FWHM  & $V_{\rm {LSR}}$ \\ 
\hline
HC$_3$N & 1.07 (0.15) & -10.93 (0.07) & 1.29 (0.18) & -10.03 (0.07) & 1.66 (0.17) & -28.53 (0.06) & 1.53 (0.33) & -50.40 (0.30) & 2.23 (0.23) & 11.53 (0.10) & 1.65 (0.26) & -39.15 (0.12) & 1.91 (0.70) & 2.46 (0.27) & 2.25 (0.22) & 6.05 (0.09) & 2.88 (0.69) & -46.80 (0.47) & 1.64	(0.17) &	-11.27 (0.08) & 2.19 (0.30) & -50.29 (0.14) \\
HC$_5$N (low) & 2.25 (0.69) & -11.25 (0.40) & 2.79 (0.78) &	-10.03 (0.56) & 2.40 (0.90) & -28.51 (0.26) & ... & ... & 2.65 (0.84) &	11.85 (0.52) & ... & ... & ... & ... & 1.76 (0.48) & 6.16 (0.37) & ... & ... & 2.67 (0.83) & -11.15 (0.92) & 1.34 (0.43) & -50.06 (0.53) \\
HC$_5$N & 2.49 (0.93) & -11.11 (0.36) & 2.11 (0.86) & -10.05 (0.34) & 2.46 (0.92) & -28.55 (0.24) & 0.82 (0.34) & -51.52 (0.45) & 2.84 (0.78) & 12.00 (0.50) & 1.02 (0.32) & -39.45 (0.26) & ... & ... & 1.51 (0.36) & 5.88 (0.45) & ... & ... & 2.53 (0.80) & -11.33 (0.45) & 1.44 (0.39) & -50.19 (0.48) \\
C$_4$H & 1.85 (0.93) & -11.02 (0.36) & 2.42 (0.68) & -9.56 (0.26) & 1.70 (0.67) & -28.42 (0.19) & 1.46 (0.39) & -50.51 (0.32) & 2.78 (0.87) & 11.63 (0.80) & 1.39 (0.39) & -38.49 (0.26) & 1.90	(0.63) & 1.95 (0.71) & ... & ... & 1.98	(0.70) & -45.77 (0.77) & 1.48 (0.36) & -11.22 (0.29) & 3.58 (0.93) & -47.77 (0.40) \\
C$_3$H & ... & ... & 2.53 (0.82) & -9.80 (0.53) & 2.48 (0.94) & -28.50 (0.28) & 1.33 (0.42) & -50.49 (0.35) & ... & ... & 1.97 (0.66) & -38.36 (0.25) & ... & ... & ... & ... & ... & ... & 1.22 (0.39) & -10.57 (0.26) & ... & ...  \\
CCS (high) & 2.61 (0.93) & -11.20 (0.40) & 2.71 (0.83) & -10.36 (0.43) & 2.49 (0.93) & -28.48 (0.27) & 1.27 (0.41) & -50.75 (0.63) & ... & ... & ... & ... & ... & ... & ... & ... & ... & ... & 1.39 (0.41)	& -11.00 (1.02) & ... & ...   \\
CCS (low) & 1.30 (0.64) & -10.86 (0.17) & 2.36 (0.85) & -9.89 (0.42) & 1.49 (0.44) & -28.31 (0.13) & 0.98 (0.20) & -50.67 (0.60) & 2.91	(0.73) & 11.57 (0.79) & 2.24 (0.75) & -38.39 (0.66) & 1.02 (0.23) & 2.73 (0.13) & 1.39 (0.40) & 5.65 (0.73) & ... & ... & 1.17 (0.22) & -11.08 (0.27) & 1.45 (0.02) & -50.29 (0.05) \\
& & & & & & & & & & & & & & & & & & & & & 1.57 (0.02)$^{(a)}$ & -47.53 (0.03)$^{(a)}$ \\
C$_3$S & 2.66 (0.85) & -11.19 (0.46) & 2.61 (0.79) & -10.53 (0.46) & 2.62 (0.88) & -28.52 (0.27) & ... & ... & ... & ... & ... & ... & ... & ... & ...& ... & ... & ... & 2.65 (0.86) & -10.74 (0.67) & ... & ... \\
$c$-C$_3$H$_2$ (high) & 2.42 (0.86) & -10.95 (0.53) & 2.46 (0.86) & -10.09 (0.25) & 2.67 (0.86) & -28.92 (0.54) & ... & ... & 2.70 (0.86) & 11.84 (0.52) & 1.32 (0.44) & -39.09 (1.01) & ... & ... & 2.68 (0.85) & 6.02 (0.54) & ... & ... & 2.75 (0.83) & -10.21 (0.52) & ... & ... \\
$c$-C$_3$H$_2$ (low) & 2.07 (0.55) & -10.99 (0.49) & 2.29 (0.87) & -10.06 (0.25) & 1.93 (0.60) & -28.66 (0.42) & 1.33 (0.44) & -50.30 (0.41) & 2.59 (0.90) & 11.88 (0.53) & 1.29 (0.46) & -38.93 (0.96) & ... & ... & 2.47 (0.93) & 5.87 (0.53) & ... & ... & 2.55 (0.88) & -10.29 (0.51) & 2.57 (0.91) & -49.75 (1.05) \\
$l$-H$_2$CCC (high) & ... & ... & 2.76 (0.79) & -9.20 (0.66) & 2.61 (0.87) & -28.51 (0.27) & 1.37 (0.44) & -51.28 (0.86) & 2.83 (0.83) & 11.96 (0.53) & ...& ...& ...& ...& ...&...&...&...& 1.89 (0.70) & -10.90 (0.55) & ... &... \\
$l$-H$_2$CCC (low) & 1.05 (0.31) & -11.20 (0.39) & 2.52 (0.93) & -9.32 (0.66) & 2.15 (0.56) & -28.48 (0.28) & 1.35 (0.43) & -51.36 (0.89) & ... & ... & 1.94 (0.69) & -38.97 (0.55) & ...&...&...&...&...&...&2.07 (0.63) & -10.96 (0.56) & ...&... \\
CH$_3$CCH & ... &... & 2.68 (0.84) & -10.14 (0.49) & 2.75 (0.81) & -28.51 (0.27) & ... & ... & 2.84 (0.76) & 11.74 (0.48) & 2.81 (0.87) & -38.09 (0.53) & ... & ...& 2.13 (0.55) & 5.99 (0.27) & ...&...& 2.71 (0.82) & -10.63 (0.79) & ... & ... \\
CH$_3$OH & ...&...&...&...&...&...&...&...&...&...&...&...&...&...&...&...& 1.56 (0.31) & -46.57 (0.24) & 1.65 (0.20) & -11.25 (0.11) & 3.16 (0.37) & -49.99 (0.15) \\
CH$_3$CHO & 2.71 (0.91)	& -11.24 (0.42) & 2.72 (0.83) & -9.95 (0.28) & 2.54 (0.91) & -28.49	(0.27) & 1.34 (0.43) & -50.32 (0.41) & 2.62 (0.90) & 11.92 (0.55) & 1.41 (0.38) & -38.42 (0.24) & ... & ... & 1.72 (0.50) & 5.96 (0.53) &...&...& 1.37 (0.42) & -10.62 (0.77) & 1.30 (0.45) & -48.82 (1.07) \\
H$_2$CCO & 2.71 (0.88) & -11.14 (0.41) & 2.63 (0.89) & -9.98 (0.27) & 2.49 (0.92) & -28.47 (0.28) & 1.34 (0.44) & -51.67 (1.09) & 2.58 (0.91) & 11.92 (0.55) & 1.32 (0.45) & -39.19 (0.80) & ...&...&...&...&...&...& 2.03 (0.63) & -11.37 (0.54) & ... & ... \\
HNCO & 2.75 (0.91) & -11.23 (0.43) & 2.69 (0.84) & -9.92 (0.26) & 2.59 (0.89) & -28.46 (0.28) & 1.40 (0.43) & -51.56 (1.11) & 2.69 (0.91) & 12.00 (0.56) & 1.72 (0.51) & -38.67 (0.41) & ... &...&1.38 (0.43) & 5.93 (0.52) & ...&...& 1.38 (0.43) & -11.11 (1.17) & ...&... \\
CH$_3$CN & 2.62 (0.91) & -11.54 (0.47) & 3.03 (0.63) & -10.03 (0.24) & 2.55 (0.92) & -28.55 (0.28) & 1.32 (0.46) & -52.16 (1.22) & 2.75 (0.86) & 11.75 (0.53) & 1.36 (0.40) & -39.00 (0.53) & ... & ... & 2.000 (0.002) & 5.05 (0.02) & 1.35 (0.42) & -46.97 (0.56) & 1.60 (0.45) & -11.04 (0.53) & ...&...  \\
 &  & & & & & & & & & & &  & & & 1.997 (0.003)$^{(a)}$ & 8.89 (0.02)$^{(a)}$  & \\
$^{13}$CS & 2.16 (0.95) & -11.47 (0.29) & 2.78 (0.88) & -9.96 (0.27) & 2.62 (0.83) & -28.50 (0.28) & 1.33 (0.45) & -51.31 (1.17) & 2.59 (0.75) & 11.62 (0.36) & 1.46 (0.36) & -38.82 (0.49) & ...&...& 2.24 (0.52) & 6.05 (0.46) & 1.93 (0.64) & -46.48 (0.27) & 2.05 (0.61) & -11.43 (0.28) & 1.96 (0.02) &-50.04 (0.03) \\
& & & & & & & & & & & & & & & & & & & & & 1.48 (0.02)$^{(a)}$ & -47.83 (0.02)$^{(a)}$ \\
C$^{34}$S & 1.26 (0.42) & -10.98 (0.14) & 1.51 (0.26) & -10.06 (0.09) & 2.22 (0.60) & -28.35 (0.18) & 1.60 (0.29) & -50.83 (0.32) & 2.48 (0.34) & 11.52 (0.14) & 1.90	(0.38) & -38.58 (0.27) & ...&...&2.48 (0.31) & 6.01 (0.18) & 2.62 (0.75) & -46.37 (0.34) & 1.63 (0.28) & -11.23 (0.12) & 1.50 (0.01) & -50.24 (0.03) \\
& & & & & & & & & & & & & & & & & & & & & 1.42 (0.01)$^{(a)}$ & -47.93 (0.01)$^{(a)}$ \\
HCS$^+$ & 2.59 (0.89) & -11.36 (0.43) & 2.63 (0.89) & -9.92 (0.25) & 2.62 (0.87) & -28.53 (0.27) & 1.41 (0.42) & -51.43	(1.05) & 2.60 (0.90) & 11.97 (0.50) & 1.37 (0.43) & -37.53	(0.27) & ... &...& 1.41	(0.41) & 5.75 (0.76) & 1.73 (0.52) & -46.45 (0.26) & 1.87 (0.46) & -10.53 (0.78) & 3.38 (1.08) & -49.40 (0.78) \\
H$_2$CS & 2.67 (0.86) & -11.20 (0.41) & 2.64 (0.90) & -9.99 (0.27) & 2.40 (0.95) & -28.49 (0.27) & 1.28	(0.46) & -51.55 (1.19) & 2.70 (1.06) & 12.02 (0.57) & 1.39 (0.40) & -39.20 (0.42) & ...&...& 1.68 (0.52) &	6.22 (0.41) & 1.27 (0.46) & -46.50 (0.28) & 1.29 (0.45) & -11.03 (0.51) & 1.30 (0.46) & -50.25 (1.27) \\
\hline
\end{tabular}
}
\tablefoot{The unit is km\,s$^{-1}$. Numbers in parentheses indicate the standard deviation error. \\(a) The 2nd velocity component.}
\end{sidewaystable*}

\end{appendix}

\end{document}